\documentclass[aps,prb,twocolumn,nofootinbib,superscriptaddress,reprint,floatfix]{revtex4}
\usepackage{color,ulem,verbatim,bbold,float,wrapfig}
\usepackage{amssymb,amsbsy,amsmath,mathrsfs}
\usepackage{graphicx}
\usepackage[utf8]{inputenc}
\usepackage[a4paper, margin = 2cm]{geometry}
\usepackage{pgf,tikz}
\usepackage{mathrsfs}
\usepackage{amssymb}
\usepackage{amsmath}
\usepackage{amsfonts}
\usepackage{subfigure}
\usepackage{hyperref}
\usepackage{braket}
\usepackage{float}
\usepackage{bbm}

\newcommand{\beq}{\begin{equation}}
\newcommand{\eeq}{\end{equation}}
\newcommand{\bea}{\begin{eqnarray}}
\newcommand{\eea}{\end{eqnarray}}

\begin{document}

\title{Emergent symmetries in prethermal phases of periodically driven quantum systems}

\author{Tista Banerjee and K. Sengupta}

\affiliation{$^{(1)}$School of Physical Sciences, Indian Association for the
Cultivation of Science, 2A and 2B Raja S. C. Mullick Road, Jadavpur Kolkata-700032, India.}

\date{\today}

\begin{abstract}

Periodically driven closed quantum systems are expected to eventually heat up to infinite temperature ; reaching a steady state described
by a circular orthogonal ensemble (COE). However, such finite driven systems may exhibit sufficiently long
prethermal regimes; their properties in these regimes are qualitatively different from that of their corresponding
infinite temperature steady states. These, often experimentally relevant, prethermal regimes host a wide range of phenomena; they may exhibit dynamical localization and freezing, host Floquet scars, display signatures of Hilbert space fragmentation, and exhibit time crystalline phases. Such phenomena are often accompanied by emergent approximate dynamical symmetries which have no analogue in equilibrium systems. In this review, we provide a pedagogical introduction to the origin and nature of these symmetries and discuss their role in shaping the prethermal phases of a class of periodically driven closed quantum systems.

\end{abstract}

\maketitle

\section{Introduction}
\label{sec:int}

The physics of closed quantum systems driven out of equilibrium has been a subject of intense research in recent years both theoretically \cite{rev1,rev2,rev3,rev4,rev5,rev6,rev7,rev8,rev9,rev10,rev11} and
experimentally using several ultracold atom \cite{rev12,exp0,exp1,exp2,exp3,exp4,exp5}, ion trap \cite{revion} and superconducting qubit \cite{revqu} platforms.
Initial endeavors in this direction constituted studies involving quench and ramp protocols \cite{rev1,rev2}.  In the former case, a parameter $\lambda$ of the Hamiltonian $H[\lambda]$ governing the dynamics system is abruptly changed at $t=0$ from $\lambda_i$ to $\lambda_f$; the subsequent evolution of the initial state (often taken as an eigenstate of $H[\lambda_i]$) is studied \cite{calabrese1,subir1,dsen0,anatoly0,sdas0}. In the latter class of drive protocols, $\lambda$ is varied
continually as a function of time with a fixed rate $\tau^{-1}$: $\lambda(t)=\lambda_i f(t/\tau)$. Usually the function $f$ is chosen such that $f(t_i/\tau)=1$ and $f(t_f/\tau)=\lambda_f/\lambda_i$ \cite{rev3,sdas1,sdas2,rajdeep1,rajdeep2,trip1}. In both cases, several quantities of the system such as excitation (defect) density $n_d$ , excess (residual) energy $Q$, correlation functions $C$, and entanglement entropies $S$ (Renyi and von-Neumann) are studied as a function of time. 

The study of such driven systems in cases where the Hamiltonian moves across or passes through a quantum critical point at $\lambda=\lambda_c$ has been of particular interest. For slow
ramp protocols, $n_d$ and $Q$ are known to scale with ramp rate $\tau^{-1}$ with universal exponents characterized by the dynamical critical exponent $z$ and the correlation length exponent
$\nu$ of the critical point. This phenomenon is known as Kibble-Zurek scaling \cite{rev1,rev2,rev3,rev4,kibble1,zurek1,anatoly1,anatoly2,dsen1,dsen2,sdas2,sdas3,sdas4,adutta1,rajdeep3}. A similar scaling phenomenon was obtained through study of fidelity $F(\delta \lambda;t) = \ln |\langle \psi(\lambda_i)|\psi(\lambda_i+\delta \lambda;t)\rangle|$ for $\delta \lambda= \lambda -\lambda_i$;  the scaling of the fidelity susceptibility  $\chi= 1-F$ has been found to be $\chi \sim |\delta \lambda|^{\alpha}$, where $\alpha$ is once again determined by the universality of the critical point \cite{anatoly0}.

The later studies of non-equilibrium dynamics concentrated on periodic drive protocols \cite{rev3,rev4,rev5,rev6,rev7,rev8,rev9,rev10,rev11}. This shift of focus occurred due to several reasons. First, the evolution operator $U$ of a periodically driven system, characterized by a drive period $T$, can be described at stroboscopic times $t= nT$ (where $n$ is an integer) by a time-independent Floquet Hamiltonian $H_F$: $U(nT,0)=\exp[-i H_F nT]$ \cite{rev8,rev9,rev10,flref1}. This provides a large simplification; several properties of the driven system can be inferred from that of a time independent Hamiltonian $H_F$. Although it is not usually feasible to have an exact analytical expression of $H_F$, several perturbative schemes exists which can provide analytical, albeit qualitative, understanding of its properties. These perturbative schemes are specially useful in the high drive frequency and/or drive amplitude regimes; examples of such schemes include Magnus expansion and Floquet perturbation theory (Magnus expansion in a rotated frame) \cite{rev11,aa1,nc1}. In contrast, for low frequencies, such techniques are unavailable for a generic non-integrable driven quantum system. The convergence of the perturbative schemes such as Magnus expansion in lower and intermediate drive frequency range is also an yet unresolved issue \cite{mori1, pre2,anatoly3,abanin1}. In the high drive frequency or amplitude regime where Magnus expansion or Floquet perturbation theory converges, the Floquet Hamiltonian can typically be approximated by a local Hamiltonian. In contrast, in the low-frequency regime, it becomes non-local. Nevertheless, even in the this regime, numerical techniques such as exact diagonalization (ED) provide us with knowledge of the Floquet spectrum specially for finite systems in $d=1$; this allows one to analyze the properties of such driven systems using a time-independent Hamiltonian.

Second, it has been shown that such driven systems often exhibit a prethermal regime before eventually heating up to infinite temperature \cite{prethermrev1,prethermrev2}. This prethermal timescale can be very large in the large drive amplitude/frequency regime; consequently, they become experimentally relevant. For example, in experiments using ultracold atom platforms \cite{rev12,exp0,exp1,exp2,exp3,exp4,exp5}, which are reasonably well isolated from the environment, have a finite system lifetime \cite{rev12}; it turns out that in certain cases the prethermal regime of the driven system can match or exceed this lifetime.  This allows one to probe the prethermal regime in these experiments. In such a prethermal regime, the driven system is well approximated by a local Floquet Hamiltonian. Moreover, it exhibits several phenomena which have no analogue in undriven quantum systems. These include dynamical freezing \cite{adas1,adas2,adas3,pekker1,deb1,uma1,camilo1,apal1,koch1,adasnew,tb1,tb2,huo1,liu1}, dynamical localization \cite{dynloc1,dynloc2,dynloc3,dynloc4,dynloc5,tanay1,fava1,rg1,galit1,martinez1,guoexp},  signatures of prethermal Hilbert-space fragmentation  \cite{sg1,sg2,xu1,zhang1}, presence of Floquet scars \cite{pretko1,bm1,mituza1,bm2,lukinsc1,papic1,liu1}, and prethermal time crystalline phases\cite{tcrev1,tcrev2,tcrev3,tcrev4,tcrev5,tcpap1,tcpap2,tcpap3}. A  detailed analysis of such phenomena has been a subject of intense research in recent years.

In this review, we shall provide a pedagogical introduction to the properties of some such prethermal phases. Often, as we shall discuss, these phases are accompanied by emergent approximate symmetries. These symmetries play a crucial role in shaping the nature of these prethermal phases. The aim of this review will be to highlight the role of these emergent symmetries and discuss their origin. In most cases which will be discussed in this review, the properties of the prethermal phases will be shown to be controlled by a local, perturbative Floquet Hamiltonian, $H_F^{(1)}$, which can be analytically obtained as the lowest order term in some properly chosen perturbation expansion. 

 It turns out that for several periodically driven systems, $H_F^{(1)}$ may respect additional symmetries or impose additional dynamical constraints at special drive frequencies or amplitudes.  These symmetries or constraints are approximate in the sense that they are not respected by the exact Floquet Hamiltonian $H_F$. However, in the large drive amplitude or frequency regime, where $H_F^{(1)}$ controls the evolution of the system up to a large prethermal time, such emergent symmetries or constraints shape the dynamics  of these driven systems.

In what follows, we shall explain this phenomenon in several contexts. The role of such emergent symmetries behind dynamical freezing in both Hermitian and non-Hermitian driven systems shall be discussed in Sec.\ \ref{frzdloc1}. This will be followed, in Sec.\ \ref{scfr} by a discussion of Floquet scars and prethermal Hilbert-space fragmentation. Next in Sec.\ \ref{tcr}, we discuss the phenomenon of prethermal time crystals . Finally, we discuss some open issues and conclude in Sec.\ \ref{diss}.

\section{Dynamical freezing and localization}
\label{frzdloc1}

In this section, we shall introduce the concept of emergent approximate symmetries using one of the simplest possible setup, namely, a periodically driven Ising model in a transverse field in $d=1$. We shall discuss, in Sec.\ \ref{frz1}, how such a driven system exhibits prethermal dynamical freezing. In Sec.\ \ref{dloc1}, we shall analyze another setup, namely, interacting 1D fermions in the presence of electric field, which exhibits dynamical localization due to a similar approximate emergent symmetry.  We note that both these phenomena involve drive-induced localization. While the former is typically used to describe localization in the Fock space, the latter refers to localization in the real space.

\subsection{Dynamical Freezing}
\label{frz1}

The $d=1$ Ising model in a transverse field is a prototypical model for studying quantum dynamics in both Hermitian and non-Hermitian contexts \cite{subirbook,lec1,nhrev}. The Hamiltonian of this model, in terms of interacting spins on a 1D chain, is given by
\begin{eqnarray}
H_{\rm Ising} &=& -J \sum_{\langle \ell j\rangle} \sigma_{\ell}^x \sigma_j^x - h_0 \sum_j \sigma_j^z, \label{isham1}
\end{eqnarray}
where $J$ is the interaction amplitude, $h_0$ is the strength of the transverse field in units of energy, $\sigma_j^{\alpha=x,z}$ denotes usual Pauli matrices on site $j$ of the chain, and $\langle \ell j\rangle$ denotes sum over sites such that the site $j$ is one of the nearest neighbors of the site $\ell$. The transverse field $h_0$ is a real number for the Hermitian model; for the non-Hermitian Ising model $h_0 = h+i \gamma$ is a complex number and $\gamma$ denotes the strength of dissipation \cite{nhrev}. We note here that the non-Hermitian transeverse field Ising model is a prototypical non-Hermitian model. It constitutes a representation of an open Ising chain in the presence of measurement in the so-called no-click limit \cite{nhrev}. In this limit, the induced non-Hermiticity in the system's Hamiltonian can be thought of as an effect of the measurement back-action on the system's degrees of freedom. Its analysis, as we discuss below, allows us to relate the steady state properties of the driven chain to emergent approximate symmetries; this feature has no counterpart in Hermitian quantum models.

We shall first consider the driven non-Hermitian model for which $h_0$ is a complex parameter. This will be followed by the results for the Hermitian case taking the limit ${\rm Im}[h_0]=0$. The drive is implemented by making $h_0$ time dependent; in what follows, we shall use 
\begin{eqnarray} 
h_0(t)= h_s+ h_1 \cos \omega_D t + i\gamma, \label{drivepr}
\end{eqnarray}
where $T=2\pi/\omega_D$ is the drive period. The central aspects of the results discussed will not depend of the precise nature of the drive protocol.

To proceed further, we follow the standard prescription and rewrite $H_{\rm Ising}$ in terms of free fermions. This is done using
the well-known Jordan-Wigner transformation given by
\begin{eqnarray}
\tau_j^{+(-)} &=& \left(\prod_{\ell=1}^{j-1} - \tau_{\ell}^z \right)
f_j^\dagger(f_j), \quad \tau_j^z= 2f_j^{\dagger} f_j -1, \label{jw1}
\end{eqnarray}
where $f_j$ denotes annihilation operator of the fermions on site
$j$. In Fourier space, these creation and annihilation operators can be written as
\begin{eqnarray}
f_j = \frac{1}{\sqrt{N}} \sum_{k\in\text{BZ}} e^{i\pi/4} e^{-ikj}\hat{f}_k,
\end{eqnarray}
where ${\rm BZ}$ indicates the first Brillouin zone $-\pi \le k \le \pi$. The phase $e^{i\pi/4}$ in the Fourier transform is introduced to make the off-diagonal part of the Fourier space Hamiltonian real for each momentum $k$.

Defining a two component fermion field in
momentum space as $\psi_k= (f_k, f_{-k}^{\dagger})^T$, where $f_k$
annihilates a fermion with momentum $k$, Eq.\ \ref{isham1} can
be written as
\begin{eqnarray}
H &=& 2 \sum_{k \in\text{BZ}/2} \psi_k^{\dagger} H_k \psi_k \label{fermham} \\
H_k &=& \tau_z (h_0(t) - \cos ka +i\gamma) + \tau_x \sin ka. \nonumber
\end{eqnarray}
Here $\vec \tau = (\tau_x, \tau_y, \tau_z)$ are standard
Pauli matrices in particle-hole space of the fermions, $J$ is set to unity,
$a$ is the lattice spacing, ${\rm BZ}/2$ denotes the positive
half of the Brillouin zone, and $\tau^{\pm} = \tau_x \pm i
\tau_y$.

To obtain the Floquet Hamiltonian of such driven system, it is useful to resort to
Floquet perturbation theory which is same, in the large drive amplitude regime, as a Magnus expansion
in the rotated frame. In this method drive amplitude is taken to be large: $h_1 \gg h_s, J, \gamma$.
One then construct a zeroth-order evolution operator from the term with the largest amplitude. For the
cosine drive protocol, this yields  
\begin{eqnarray}
U_{0k}(t,0) &=&  e^{-2 i h_1 \sin \omega_d t \tau_z /(\hbar \omega_D)}, \label{zeroU}
\end{eqnarray}
for a momentum mode $k$. We note that $U_{0k}$ has a single Pauli matrix which allows one to take care of the time ordering
easily. Furthermore, $U_{0k}(T,0)= I$ (where $I$ denotes the identity matrix) leading to $H_{Fk}^{(0)}=0$
for all $k$. 

Next one obtains the higher order terms in $U(T,0)$. The first order perturbative correction to the evolution
operators is obtained as \cite{tb1,tb2,rev11,aa1,nc1}  
\begin{eqnarray}
U_{1k}(T,0) &=& \left(\frac{-i}{\hbar}\right) \int_0^T dt  U_{0k}^{\dagger} (t,0)  H'_k  U_{0k}(t,0) \nonumber \\
H'_k &=&  2( \tau_z (h_s-\cos ka + i\gamma) + \tau_x\sin ka) \label{u1eq} 
\end{eqnarray}
This expression can be evaluated in a straightforward manner and one obtains $H_{Fk}^{(1)}= i\hbar U_{1k}(T,0)/T$ 
given by \cite{tb1,tb2}
\begin{eqnarray}
H_{Fk}^{(1)} &=& 2  \Big[\tau_z(h_s - \cos ka + i\gamma) \nonumber\\
&& + \tau_x  J_0(\mu) \sin ka \Big],  \nonumber\\
H_F^{(1)} &=& \sum_k \psi_k^{\dagger} H_{F k}^{(1)} \psi_k. \label{hf1exp}
\end{eqnarray}
Here $J_p$ denotes the $p^{\rm th}$ order Bessel function of the first kind and $\mu=4 h_1/(\hbar \omega_d)$.
The first term in Eq.\ \ref{hf1exp} follows from the fact that the first term of $H'_k$ commutes with $U_{0k}$.
For obtaining the second term, one needs to note that the action of $\tau_x$ on an eigenstate of $\tau_z$ simply flips
a pseudospin. The details of the relevant calculation is starightforward and presented in Refs.\ \onlinecite{tb1,tb2}.
The second order term in the perturbative expansion can also be obtained in a similar manner. A somewhat
cumbersome calculation, outlined in Refs.\ \onlinecite{tb1,tb2}, leads to
\begin{eqnarray}
&& H_{F k}^{(2)} = \Big[  - 4 \tau_z \sin^2 ka \sum_{n=0}^{\infty} \frac{
J_0(\mu) J_{2n+1}(\mu)}{(2n+1)\hbar \omega_D} \nonumber\\
&& + \tau_x \sin ka (h_s-\cos ka +i \gamma)
\sum_{n=0}^{\infty} \frac{4 J_{2n+1}(\mu)}{(2n+1) \hbar \omega_D} \Big]. \nonumber \\
&& H_F^{(2)} = \sum_{k \in\text{BZ}/2} \psi_k^{\dagger} H_{F k}^{(2)} \psi_k. \label{hf2exp}
\end{eqnarray}
We note that $H_{Fk}^{(2)}$ is suppressed by a factor of $1/\omega_D$ compared to $H_{F k}^{(1)}$ indicating the efficacy of the
perturbative expansion at high drive frequencies.

In what follows we are going to analyze the properties of the system using the expressions of $H_{Fk}^{(1)}$ and $H_{Fk}^{(2)}$. To this end, 
we first set $\gamma=0$ and focus on the Hermitian Ising chain. From $H_{Fk}^{(1)}$ we note that at special frequencies  $\omega_D=\omega_m^{\ast}$, for which
$J_0(\mu)=0$; this leads to
\begin{eqnarray}
\frac{4 h_1}{\hbar \omega_{m} ^{\ast}} = \beta_m,
\label{cond1}
\end{eqnarray}
where $\beta_m$ denotes $m^{\rm th}$ zero of $J_0(\mu)$. At these drive frequencies, the second term in Eq.\ \ref{hf1exp} vanishes; consequently  $H_{Fk}^{(1)}$ commutes with $\tau_z$. This leads to an additional approximate emergent symmetry resulting in conservation of the fermion density $n_k = \psi_k^{\dagger} \tau_z \psi_k$ (which is equivalent to conservation of the magnetization $M_z = \sum_j \sigma_j^z$ in the original spin model) at stroboscopic times $t= nT$ where $n$ is an integer. This emergent symmetry is approximate since it is broken by $H_{Fk}^{(2)}$; this is easily seen from the second term of Eq.\ \ref{hf2exp} which does not commute with $\tau_z$ at these frequencies. We note here that such emergent symmetries occur for other drive protocols; however, the condition given by Eq.\ \ref{cond1} for their occurrence changes. For example, a square-pulse drive protocol, which has been studied extensively in the literature in this context \cite{adasnew, tb1,tb2,sg1}, leads to a condition 
$2 h_1/(\hbar \omega_m^{\ast})= m$ for integer $m$.

\begin{figure}
\rotatebox{0}{\includegraphics*[width= 0.49 \linewidth]{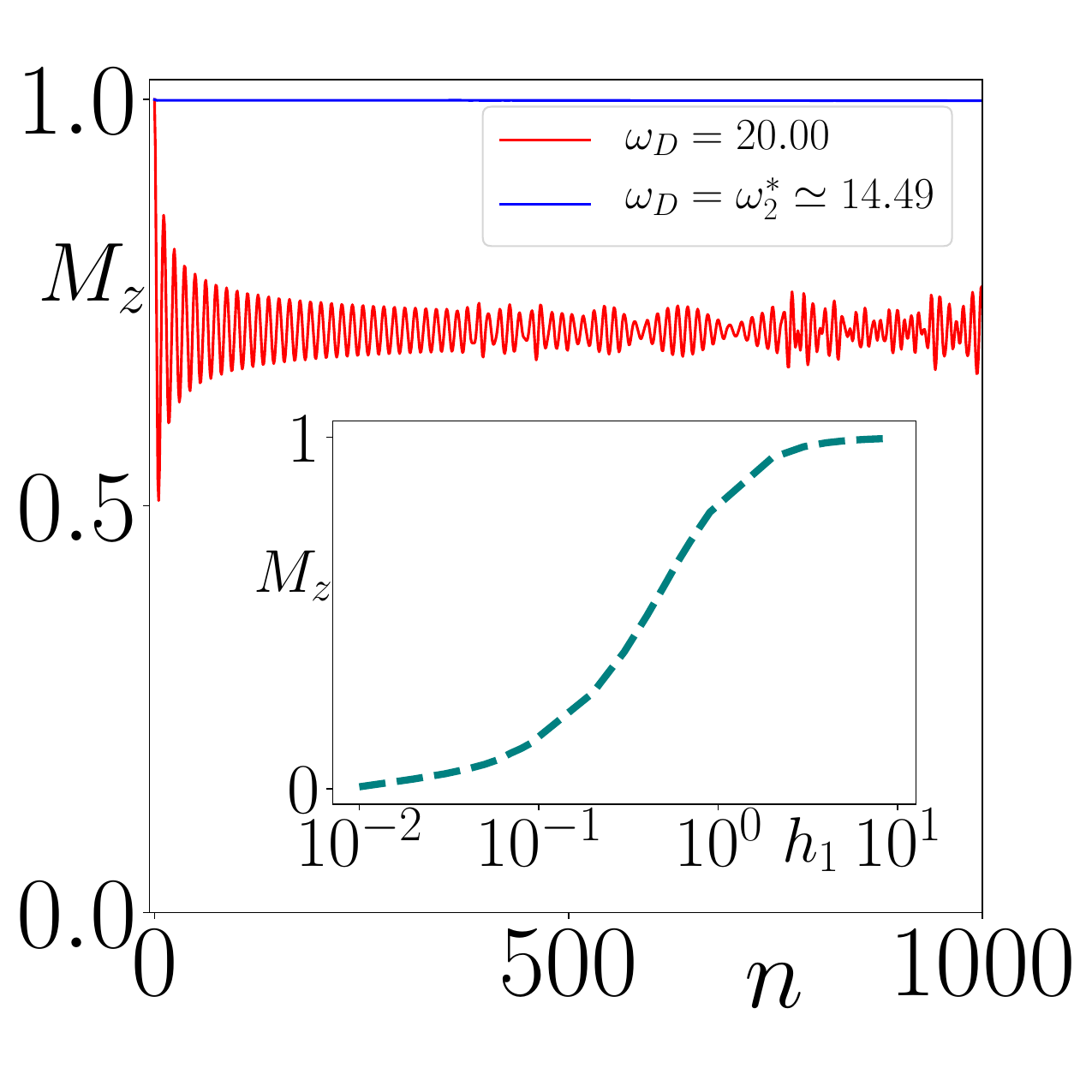}}
\rotatebox{0}{\includegraphics*[width= 0.49 \linewidth]{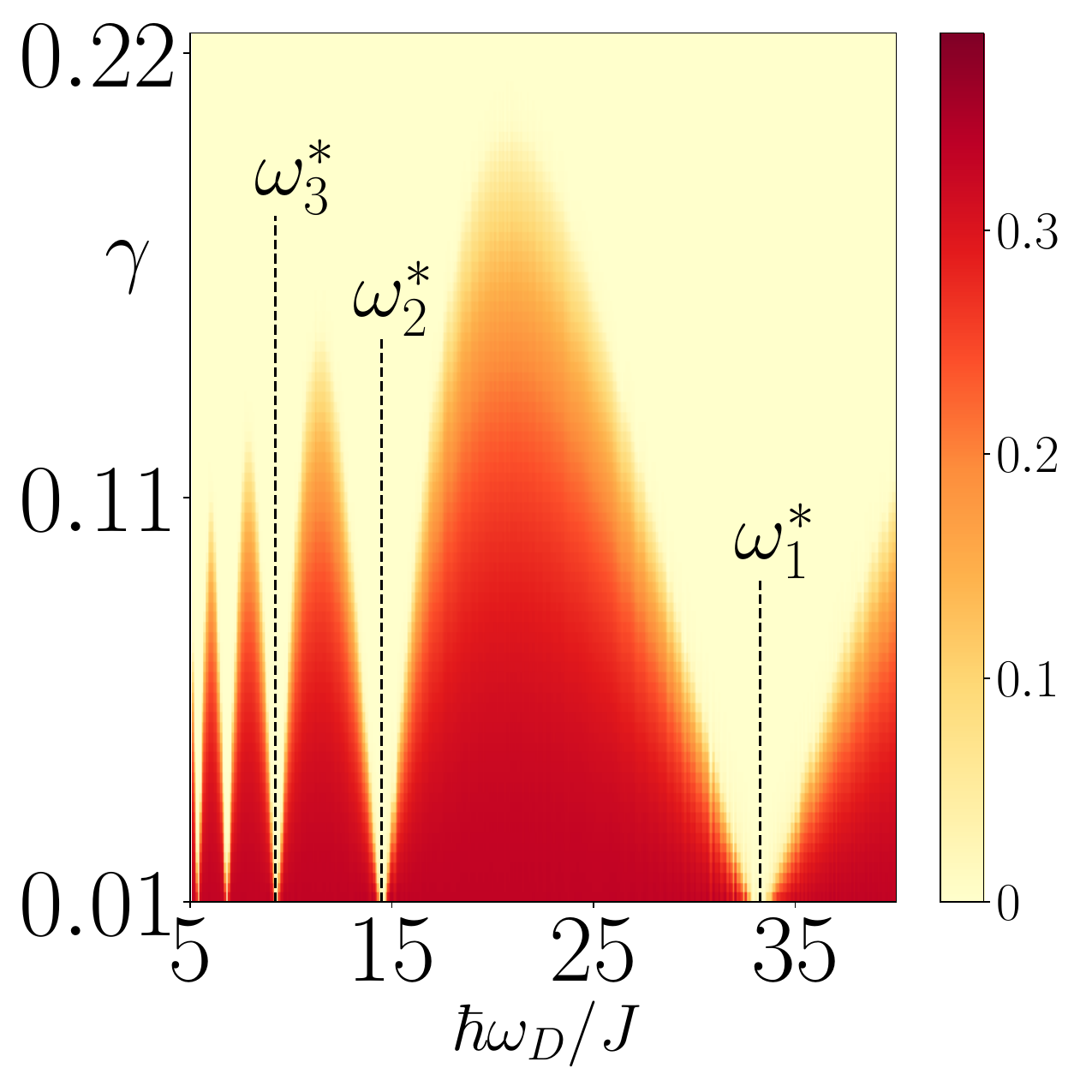}}
\caption{ Left Panel: Plot of $M_z$ as a function of number of drive cycles $n$ showing dynamical freezing
at special drive frequency $\omega_D=\omega_2^{\ast}= 4 h_1/\beta_2$ and evolution to its
steady state value away from the special frequency (red curve). The inset shows $M_z$ averaged over $n_0=200$
cycles (from $n= 1800$ to $n=2000$) as a function of $h_1$ at $\omega_D=\omega_1^{\ast}$ indicating loss of
freezing at low drive amplitude. For all plots $h_1 = 20$, $h_s = 0.1$,
$L=500$, and $J =1$.  Right Panel: Plot of $\alpha$ (where $ S_{L/2} \sim \alpha \ln L$)
as a function of $\omega_D$ and $\gamma$ for a cosine drive protocol. The plot, adapted from Ref.\ \onlinecite{tb2}, 
clearly indicates that $\alpha$ vanishes at special drive frequencies $\omega_n^{\ast}$ indicated in the figure.
For all plots $h_1=20 J$, $h_s=0.1$, $J=1$ and $L =1000$. See text for
details. \label{figsec2en}}
\end{figure}

 The consequence of such an emergent symmetry is dynamical freezing of the magnetization $M_z$ (or equivalently $\sum_k \langle n_k\rangle$) at stroboscopic times $t= nT$ up to a prethermal timescale where $H_F^{(1)}$ dominates the dynamics (left panel of Fig.\ \ref{figsec2en}). This phenomenon was first noted in Ref.\ \onlinecite{adas1}; there are several subsequent works on this phenomenon which confirmed this behavior \cite{adas2,adas3,pekker1,adasnew} (left panel of Fig.\ \ref{figsec2en}). Such a freezing is approximate; the dynamics of $M_z$ is eventually restored when effects of $H_F^{(2)}$ and other higher order terms becomes important. 

 A precise understanding of this prethermal timescale, for Magnus expansion, is given in Ref.\ \onlinecite{mori1}. It demonstrates that the prethermal timescale after which the Magnus expansion breaks down and thermalization sets in is ${\rm O}(\exp[c\omega_D])$, where $c$ is typically a number determined by system timescale (for the present system, $c\sim J^{-1}$). Thus at low or intermediate frequencies, where $\omega_D c \le 1$, thermalization sets in within the first few drive cycle. In contrast, for $\omega_D c\gg 1$, the prethermal timescale is exponentially large. This is shown in the inset of the left panel of Fig.\ \ref{figsec2en} which plots the time-averaged value of $M_z$ at large $n$ as a function of $h_1$ at $\omega_D=\omega_2^{\ast}$. The plot clearly shows that the system crosses over to a freezing regime at large $\omega_D$.

Numerical analysis for finite chains at very large drive amplitude with $\omega_D=\omega_1^{\ast}$ indicates that the freezing timescale is not reached within $10^5$ drive cycles. This makes the freezing almost exact from an experimental point of view since it may easily exceed the system lifetime in an ultracold platform. Such an exponential dependence of the prethermal timescale on the drive frequency or amplitude in a similar setup has been numerically verified in Ref.\ \onlinecite{sg1}; this shall be discussed in Sec.\ \ref{scfr} in more details. We note here that analogous phenomenon has been theoretically studied in the context of driven chaotic quantum dots \cite{deb1}, driven tilted Bose-Hubbard model \cite{uma1}, non-equilibrium Ising phase transitions \cite{camilo1}, and spin-half staggered Heisenberg model in a transverse field \cite{apal1}. It has also been experimentally investigated using ultracold $^{87}{\rm Rb}$ atoms \cite{koch1} and nuclear spins where measurements were carried out using nuclear magnetic resonance (NMR) techniques \cite{adas3}.

In the non-Hermitian context ($\gamma \ne 0$), such an emergent symmetry influences the steady state behavior. To understand this we note that in these systems, the time taken to reach the steady state depends on $\gamma$; consequently, as seen in Refs.\ \onlinecite{tb1,tb2}, at large drive frequencies, the steady state may be reached within this prethermal timescale where the properties of $H_F^{(1)}$ controls the steady state behavior. A detailed study of correlation functions \cite{tb1} and entanglement dynamics \cite{tb2} clearly indicates that this is the case. Interestingly, the behavior of half-chain entanglement entropy, $S_{L/2}$, also changes due to such emergent conservation. As shown in Ref.\ \onlinecite{tb2}, for small $\gamma$, $S_{L/2} \sim \alpha \ln L$,  while for large $\gamma$, $S_{L/2}$ shows an area-law behavior and thus become independent of $L$. In between, there is a transition, termed as entanglement transition in Ref.\ \onlinecite{tb2}, which occurs at critical $\gamma_c$ where $\alpha\to 0$. We note that such a transition is characterized by the drive frequency in contrast to the more well-known measurement-rate induced entanglement transition studied, for example, in Refs.\ \cite{fisher1,nahum1,huse1}. It has been shown in Ref.\ \onlinecite{tb2} that the entanglement for such a driven chain, exhibits a logarithmic dependence for $\gamma \le \gamma_c$, where $\gamma_c \sim J_0(\mu)$; consequently, $\gamma_c$ vanishes at the special drive frequencies. Thus, at the special drive frequencies the entanglement always follow an area law; this can be seen from the right panel of Fig.\ \ref{figsec2en} where the coefficient $\alpha$ is plotted as a function of $\gamma$ and $\omega_D$. We note that freezing, with focus on information scrambling, has also been studied for periodically kicked non-Hermitian rotors \cite{huo1,liu1}.

Before ending this section, we note that for Hermitian systems in all the above-mentioned cases, the prethermal timescale over which freezing takes place is finite, although it is exponentially large in the large drive amplitude regime. In contrast, there have been recent claims regarding exact freezing in Ref.\ \onlinecite{adasnew}; the exact nature of such freezing, if it exists, remains to be rigorously proven. Also, we note here that although a single parameter drive usually does not lead to exact freezing, carefully chosen protocol may lead to freezing in non-interacting free fermion systems \cite{rudner1}. For interacting systems, a two-tone protocol can lead to exact dynamical freezing in interacting many-body systems via realization of exact Floquet flat bands; such freezing necessarily requires two parameters driven with frequencies having a specific ratio and can not be understood from the perspective of an emergent symmetry \cite{tb0,skar1,skar2}.

\subsection{Dynamical localization}
\label{dloc1}

The phenomenon of dynamical localization constitutes freezing of transport in a periodically driven quantum system \cite{dynloc1, dynloc2,dynloc3,dynloc4,dynloc5}.  This phenomenon has been studied in the context of interacting models with on-site periodic potentials \cite{dynloc4}, driven hardcore bosons in their superfluid phases \cite{dynloc5}, periodically kicked three-band models \cite{tanay1}, driven Bose-Hubbard model \cite{fava1}, interacting, driven fermions \cite{rg1}, periodically kicked rotors \cite{galit1}, in driven two-level systems studied using a Green function approach \cite{martinez1}, and to implement coherent destruction of tunneling between particles on a lattice \cite{tun1}. Experimentally, there has been a
recent report of observing many-body dynamical localization \cite{guoexp} as well as several works on dynamical localization in ultracold atoms \cite{ari1} 
It has some similarity with dynamic freezing since it constitutes drive induced freezing that leads to localization. 

As mentioned above, dynamical localization has been studied in many different contexts. 
In what follows, we shall understand the effect of approximate emergent symmetry behind it by using a simple model, namely,
interacting fermions on a 1D lattice in the presence of a time-dependent electric field \cite{dynloc1}. The Hamiltonian of such a model is given by
\begin{eqnarray}
H_{\rm ferm} &=& -\sum_{j \sigma=\uparrow, \downarrow} \left(J' c_{j, \sigma}^{\dagger} c_{j+1, \sigma}+ {\rm h.c.} \right) \nonumber\\
&& + U_0 \sum_j \hat n_{j \uparrow} \hat n_{j \downarrow} - eE(t) \sum_{j , \sigma= \uparrow, \downarrow} j \hat n _{j \sigma},
\label{hamdloc1}
\end{eqnarray}
where $c_{j \sigma}$ denotes annihilation operator for fermions with spin $\sigma=\uparrow,\downarrow$ on site $j$ of the lattice $E(t)= E_0 \cos \omega_D t$ is the applied electric field, $U_0$ is the on-site interaction strength,  and $\hat n_{j \sigma} = c_{j \sigma}^{\dagger} c_{j \sigma}$ is the density of spin $\sigma$ fermions at site $j$.

In the non-interacting limit where $U_0=0$, $H_{\rm ferm}$ can be most easily analyzed using the equation of motion method outlined in Ref.\ \onlinecite{dynloc6}. In this limit, the Heisenberg equation of motion for $c_j$ is given by
\begin{eqnarray}
i \partial_t c_{j, \sigma} &=& -J' ( c_{j-1, \sigma}+ c_{j+1, \sigma}) - e E(t) j c_{j, \sigma}.  \label{eom1}
\end{eqnarray}
To solve this equation, one moves to a rotating frame, using a transformation $ \tilde c_{j, \sigma}(t) = U(t,0) c_{j, \sigma}(t)$, where
\begin{eqnarray}
U(t,0) &=& \exp[i e \sum_j j \hat n_{j \sigma}  \int^t_0 dt' E(t')/\hbar] \nonumber\\
&=& \exp[-i e \sum_j j \hat n_{j \sigma}  {\mathcal A}(t) /\hbar],  \label{trans}
\end{eqnarray}
and ${\mathcal A}(t)$ is the vector potential with $\dot {\mathcal A}(t) = -E(t)$. In the transformed frame, the equation of motion $\tilde c_{j \sigma}$ admits an easy solution
\begin{eqnarray}
\tilde c_{j \sigma}^{k} (t)= \exp[i ( k j- f(k,t))/\hbar] \tilde c^{k}_{\sigma}(0),
\end{eqnarray}
where $\dot f= -2 J' \cos(k + {\mathcal A}(t))$, and $\tilde c^k_{\sigma}$ is the annihilation operator of a fermion in the rotated basis with spin $\sigma$ at momentum $k$.  Using this
solution, one can find that the fluctuation of fermion occupation $\langle n^2\rangle = \langle \sum_j \hat n_j^2 \rangle$ at half-filling to be \cite{dynloc1}
\begin{eqnarray}
\langle n^2 \rangle &=& 2 J' \left[ ( J_0(x) t + \mu(t))^2 + \nu(t)^2 \right]/\hbar  \label{flden}\\
\mu(t) [ \nu (t)] &=& \sum_{n \ne 0} \frac{J_n(x) \sin n \omega_D t [(1-\cos n \omega_D t)]}{n \omega_D}.  \nonumber
\end{eqnarray}
where $x= eE_0/(\hbar \omega_D)$ and we have omitted the spin index for simplicity since Eq.\ \ref{flden} holds for both spins. 

\begin{figure}
\rotatebox{0}{\includegraphics*[width= 0.98 \linewidth]{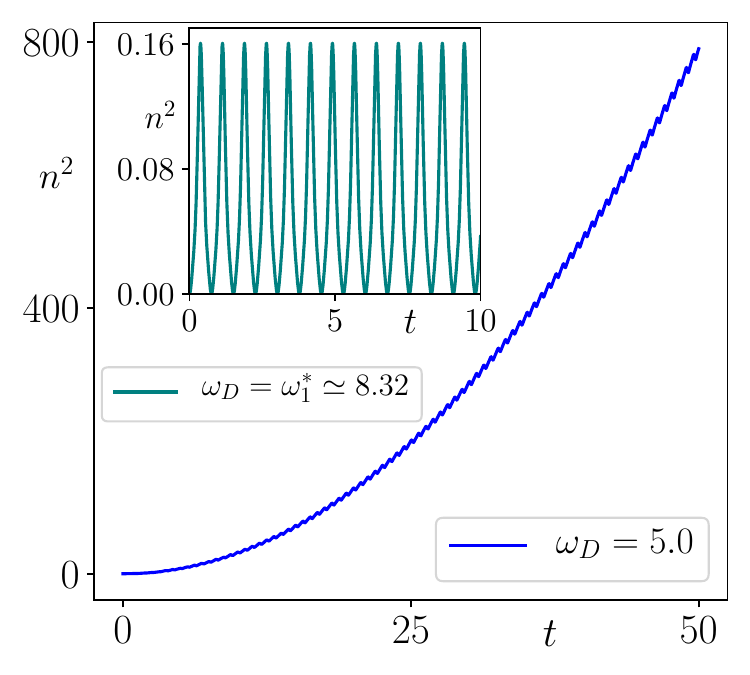}}
\caption{Plot of $\langle n^2\rangle$ as a function of $t$ away from the special drive frequency
showing its unbounded nature indicating lack of localization. The inset shows behavior of $\langle n^2\rangle$ as a function of $t$
at the special frequency $\omega_1^{\ast}$. $\langle n^2\rangle$ is bounded showing dynamical localization; moreover, it
returns to its initial value at stroboscopic times $t = 2\pi p/\omega_D$ (for integer $p$).
For all plots $J'=1= E_0$ and $L =1000$. See text for details. \label{figsec2dloc}}

\end{figure}

From Eq.\ \ref{flden}, we note that at special frequencies given by $x= \beta_m$, where $J_0(\beta_m)=0$, the motion of the fermions are bounded since $\langle n^2\rangle$ is a bounded function of $t$. This can be seen from expressions of $\mu(t)$ and $\nu(t)$ in Eq.\ \ref{flden}. Moreover, at stroboscopic times $t= 2 \pi p/\omega_D$, where $p$ is an integer, $\langle n^2\rangle=0$. At these times, the system, if driven at special frequencies, returns to its initial state. These features are shown in Fig.\ \ref{figsec2dloc} and constitute dynamical localization of non-interacting fermions.

The stability of such dynamical localization in the presence of interaction requires an analysis of the interacting Hamiltonian. Such an analysis, carried out in Ref.\ \onlinecite{dynloc1}, is most easily done by introducing the concept of dressed fermions. These dressed fermions are 
bound state of a bare fermion with $\ell$ photons. The annihilation operator for the dressed fermions is given by $f_{j \sigma \ell}$; it is 
related to $c_{j \sigma}$ by 
\begin{eqnarray}
c_{j \sigma} (t) &=& e^{- i \epsilon t} \sum_{\ell} f_{j \sigma \ell} e^{-i \ell \omega_D t}. \label{dress1}
\end{eqnarray}
The density of such composite objects can be represented by density operators $\hat n_{j \sigma \ell} = \sum_{\ell'} f_{j \sigma \ell'}^{\dagger} f_{j \sigma \ell+\ell'}$. In terms of these composite operators, the Floquet Hamiltonian of the system has been found to be \cite{dynloc1}
\begin{eqnarray}
H_F &=& -J' \sum_{j, \sigma, \ell, \ell'}  \left( J_{\ell-\ell'}(x) f_{j \sigma \ell}^{\dagger} f_{j+1 \sigma \ell'} + {\rm h.c.}\right) \nonumber\\
&& + U_0 \sum_{j \ell}  \hat n_{j \uparrow \ell} \, \hat n_{j \downarrow  \ell} + \sum_{j \sigma \ell} \ell \omega_D  \hat n_{j \sigma \ell}.
\end{eqnarray}

To understand the nature and role of the approximate emergent symmetry in this context, one notes that $H_F$ has two energy scales, $x_1= U_0/J'$ and $x_2= \hbar \omega_D/J'$. In the limit when $x_2$ is the largest energy scale, only terms with $\ell=\ell'$ in $H_F$ contribute. In this regime, for a drive frequency which satisfies $e E_0/(\hbar \omega_D)= \beta_m$, one finds that there is no hopping of undressed fermions.
We note that here we are restricted to large drive amplitude limit since the ratio $eE_0/(\hbar \omega_D)$ is fixed.

In contrast, the dressed fermions hopping is finite, but suppressed by a factor of $1/\omega_D$, similar to the suppression of $H_F^{(2)}$ in the context of dynamic freezing (Eq.\ \ref{hf2exp}). Ignoring the latter terms at high drive frequencies, one finds that up to a large prethermal timescale, the bare fermion density, $c_{j \sigma}^{\dagger} c_{j \sigma}$, at any site $j$ commutes with the first order Floquet Hamiltonian. This indicates localization of undressed fermions in the prethermal regime. 

This prethermal dynamical localization at stroboscopic times is a consequence of the approximate
emergent symmetry which conserves local fermion density. However, this localization is approximate at any finite drive frequency. This is due to the interaction between the fermions which makes it impossible to achieve full localizations and leads to the presence of mobile dressed fermions. This feature has been alluded to in different contexts in the literature where effect of interaction on dynamical localization have been studied \cite{dynloc2,dynloc3,dynloc4,dynloc5}.

\section{Prethermal Hilbert-space fragmentation and Floquet scars}
\label{scfr}

In this section, we shall first, in Sec.\ \ref{phsf}, discuss the realization of prethermal Hilbert-space fragmentation in driven system and the role of an emergent approximate symmetry behind it. This will be followed by a discussion of prethermal realization of quantum many-body scars (QMBS) in Sec.\ \ref{qscars}.

\subsection{Prethermal Hilbert-space fragmentation}
\label{phsf}

The phenomenon of strong Hilbert-space fragmentation was first pointed out for fractonic models \cite{hsf1,hsf2} and interacting spin model on a chain \cite{hsf3} in the context of time-independent Hamiltonians. Since then, this phenomenon has been studied using several models in many different contexts \cite{hsf4,hsf5,hsf6,hsf7,hsf8,hsf9,hsf10,hsf11,hsf12,hsf13,hsf14,hsf15,hsf16,hsf17,hsf18,hsfrev,hsfexp,as1,frank2,ac1,sid1,gia1,nh1,open1}. In earlier studies \cite{hsf1,hsf2,hsf3,hsf4}, such fragmentation was shown to be always accompanied by two global symmetries; in most cases these symmetries correspond to number and dipole conservation. 

The phenomenon of Hilbert-space fragmentation constitutes fragmentation of the system Hilbert space into an exponentially large number of dynamically disconnected sectors in the computational (often taken to be Fock) basis. In other words, for these systems it is possible to construct dynamically disconnected distinct Krylov subspaces, ${\mathcal K}_j$, such that the total Hilbert space dimension ${\mathcal D}_t = \bigoplus {\mathcal{K}_j}$. Such spaces can be span by repeated application of $H$ of an initial state $|\psi_j\rangle$; $\mathcal{K}_j = \mathcal{\text{span}} \{ \ket{\Psi_j},\text{H}\ket{\psi_j},\text{H}^2\ket{\psi_j},...\}$. 

For strong Hilbert-space fragmentation, there are an exponentially large number of such Krylov subspaces. The Hilbert space dimension (HSD) of the largest fragment (or equivalently largest Krylov subspace), ${\mathcal D}_L$, turns out to be exponentially smaller compared to the total HSD ${\mathcal D}_t$: ${\mathcal D}_L/{\mathcal D}_t \sim \exp[-c L]$, where $L$ is the system size in units of lattice spacing and $c$ is a ${\rm O}(1)$ non-universal constant. This leads to an ${\rm O}(e^L)$ fragments and distinguishes this phenomenon from separation of the Hilbert space into dynamically disconnected symmetry sectors; the number of the latter sectors is at most ${\rm O}(L)$. Such a fragmentation naturally leads to loss of ergodicity and constitutes a strong violation of eigenstate thermalization hypothesis (ETH) \cite{ethref1,ethref2,ethref3}. This phenomenon is often termed as {\it strong Hilbert space fragmentation}; for the rest of this section, we shall omit the qualifier "strong" for brevity.  

It turns out that there are models which exhibit Hilbert-space fragmentation without the presence of two or more global conserved quantities; such conservations could be emergent within a fragmented sector \cite{hsf6,hsf9,hsf13,hsf14,hsf17,as1}. Moreover, fragmentation need not only occur in the classical Fock basis, but can also happen in an entangled basis. This phenomenon was first pointed out as secondary fragmentation in Ref.\ \onlinecite{as1}; it has been later dubbed as quantum fragmentation and discussed in several works \cite{hsf10,hsf11}. Most of the models discussed for studying Hilbert-space fragmentation are one-dimensional \cite{hsfrev}; however, more recently some higher-dimensional models exhibiting Hilbert-space fragmentation has been analyzed in details \cite{frank2,ac1,sid1}. The effect of the range of interaction on Hilbert-space fragmentation has also been studied using Fredkin chains with longer range interactions \cite{gia1}. The presence of fragmentation in non-hermitian and open quantum models has also been investigated \cite{nh1,open1}.

The prethermal Floquet version of this phenomenon has been discussed in Refs.\ \onlinecite{sg1,sg2,xu1,zhang1}. Here we shall closely follow the model used in Refs.\ \onlinecite{sg1,sg2}, namely, a driven spinless fermionic chain with nearest and next-neighbor density-density interaction and nearest-neighbor hopping. The driven Hamiltonian, introduced in Ref.\ \onlinecite{sg1}, is given by
$H(t)= H_0(t) +H_1$, where
\begin{eqnarray}
H_0(t) &=&  V(t) \sum_{j=1..L} \hat n_j \hat n_{j+1}, \label{fermham} \\
H_1 &=& \sum_{j=1..L} -J (c_j^{\dagger} c_{j+1} + {\rm h.c.}) \nonumber\\
&& + \hat n_j (V_0 \hat n_{j+1} + V_2 \hat n_{j+2}). \nonumber
\end{eqnarray}
Here $c_j$ denotes the fermion annihilation operator for the site
$j$ of the chain, $\hat n_j= c_j^{\dagger} c_j$ is the fermion
density operator, $V_0+V(t)$ and $V_2$ are the strengths of nearest-
and next-nearest neighbor interactions respectively. 

The drive is implemented using
a square-pulse protocol: $V(t)= -(+) V_1$ for $t
\le (>) T/2$, where $T= 2\pi/\omega_D$ is the time period of the drive and $\omega_D$ is the drive frequency.
We note that $H(t)$ conserves fermion density: $[ H(t), N]=0$ where $N=\sum_j \hat n_j/L$ and $L$ is chain length.
\begin{figure}
\rotatebox{0}{\includegraphics*[width= 0.49 \linewidth]{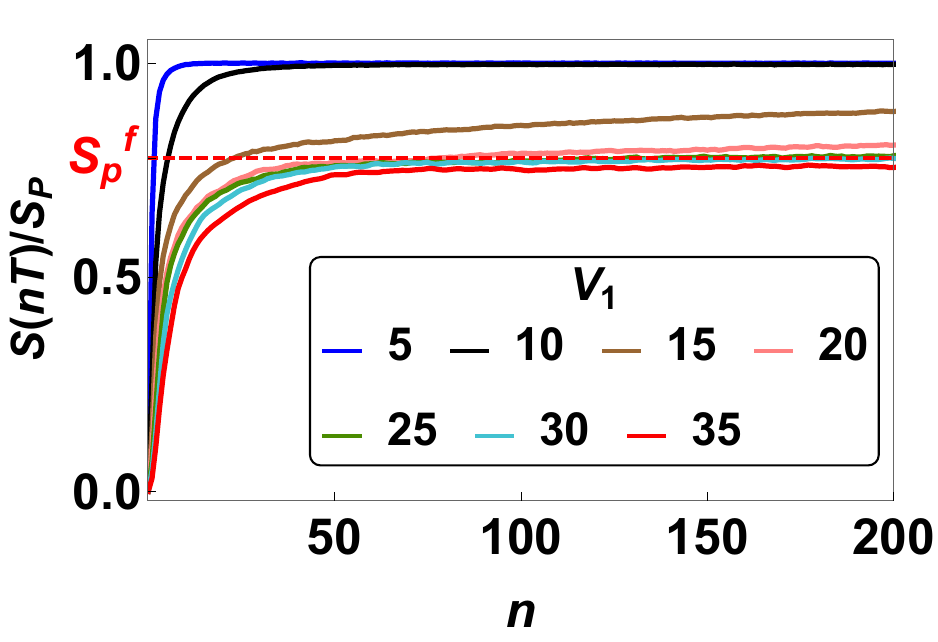}}
\rotatebox{0}{\includegraphics*[width= 0.49 \linewidth]{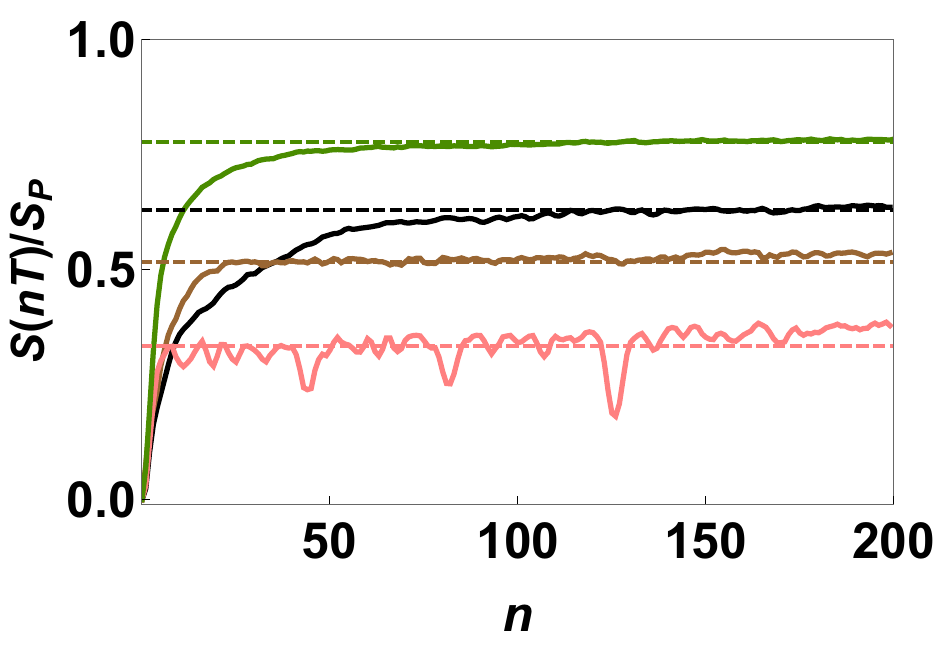}}
\rotatebox{0}{\includegraphics*[width= 0.49 \linewidth]{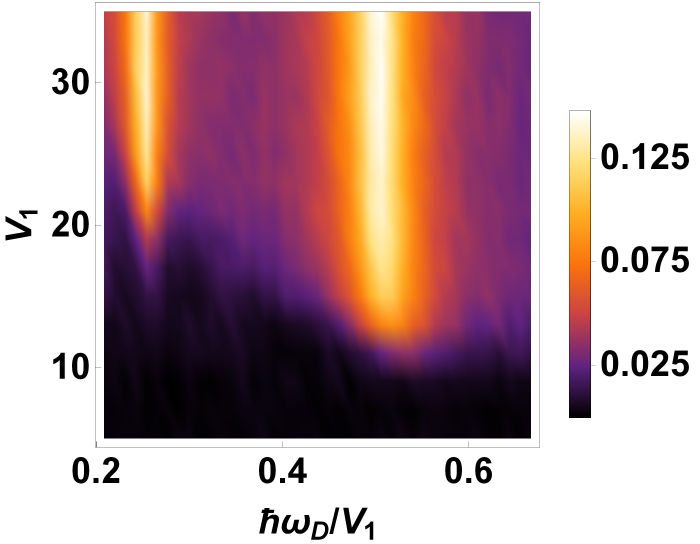}}
\rotatebox{0}{\includegraphics*[width= 0.49 \linewidth]{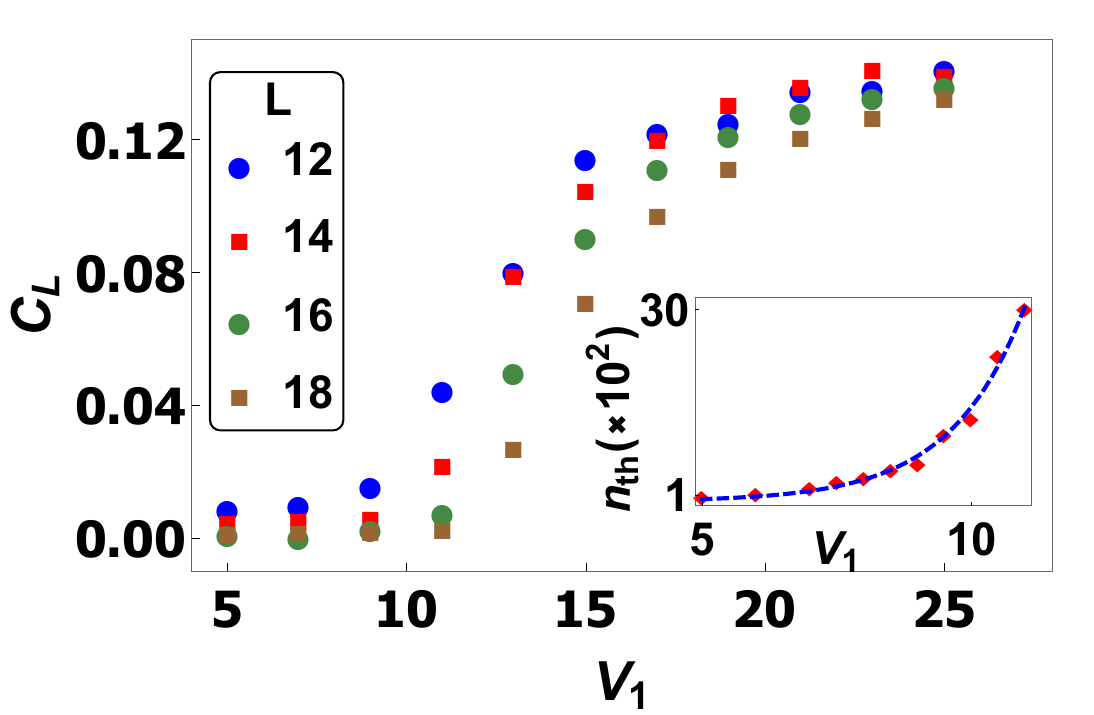}}
\caption{ Top left: Plot of half-chain entanglement entropy $S(nT)/S_p$ as a function of $n$ and $\hbar \omega_D = V_1/2$ and
several $V_1$  starting from a Fock state belonging to a fragment with ${\rm HSD}=1008$ and $S_p^f = 0.8 S_p$ (red dotted line).
$S$ saturates to $S_p(S_p^f)$ at low(high) $V_1$. Top right: Plot of $S(nT)/S_p$ for random initial Fock states saturating to different
$\eta$. Bottom left:  Plot of $C_L(nT)$ for $n=5000$ as a function of $\hbar \omega_D/V_1$ showing non-zero long time value of $C_L$ at special frequencies
where $\hbar \omega_D/V_1= 1/2, 1/4$. Bottom Right:  Plot of $C_L(nT)$ for $n = 5000$ as a
function of $V_1$  at $\hbar \omega_D= V_1/2$ for several chain length $L$ showing a clear crossover
to a prethermal Hilbert-space fragmentation regime at large $V_1$. The inset shows the
number of cycles, $n^{\rm th}$, required for $C_L(nT)$ to reach its Floquet ETH
predicted infinite-temperature value for $L = 16$ at $\hbar \omega_D= V_1/2$; $n^{\rm th}$ shows
exponential scaling with $V_1$ at large $V_1$.  All figures are obtained from Ref.\ \onlinecite{sg1}.} \label{hsdfig}
\end{figure}
In the large drive amplitude regime, the leading term of the Floquet Hamiltonian can be computed within
FPT and yields \cite{sg1,sg2}
\begin{eqnarray}
H_F^{(1)} &=& \sum_{j=1..L} \hat n_j ( V_0 \hat n_{j+1} +V_2 \hat
n_{j+2})
\label{fl1} \\
&& -J \sum_{j} [ (1-\hat A_j^2) + \alpha(\gamma_1) \hat A_j^2]
c_j^{\dagger} c_{j+1} +{\rm h.c.},\nonumber
\end{eqnarray}
where $\hat A_j =(\hat n_{j+2}- \hat n_{j-1})$, $\gamma_1= V_1
T/(4\hbar)$ and $\alpha(\gamma_1)= \gamma_1^{-1} \sin \gamma_1 \exp[i
\gamma_1 \hat A_j]$. 

For $\gamma_1=n \pi$, or $V_1=2 n \hbar \omega_n^{\ast}$, $\alpha(\gamma_1)=0$.
At these special frequencies, $H_F^{(1)}$ represents a model of interacting 1D spinless
fermions with constrained hopping; such a model, in equilibrium, leads to strong Hilbert-space fragmentation \cite{hsf4}.
However, higher order terms in $H_F$ (Ref.\ \cite{sg2}) do not respect such constraints. Most importantly, the
presence of such constraints leads to two approximate emergent conservations; $[H_F^{(1)}, n_d]=0$ and $[H_F^{(1)}, n_d^s]=0$,
where $n_d= \sum_j \hat n_j \hat n_{j+1}$ and $n_d^s= \sum_j (-1)^j \hat n_j \hat n_{j+1}$. Note that these conservation 
laws are approximate as higher order terms in $H_F$ do not respect them. These approximate emergent
conservation laws are crucial in obtaining a prethermal Hilbert-space fragmentation phase at special frequencies.

The signature for such Hilbert-space fragmentation becomes apparent in several quantities. For example, one expects the entanglement entropy for a generic initial state of an ergodic driven system to eventually saturate to its Page value $S_p$. This can be thought as the average entropy for a generic ergodic many-body system; it is also the typical entropy for large enough systems and depends on the HSD of the system (see Refs.\ \onlinecite{pageref, dona1} for details). However, at these special frequencies and for high drive amplitude, the half-chain entanglement entropy $S(nT)$, of an initial Fock state after $n$ drive cycles does not saturate to the corresponding Page value $S_p$  as expected in a ergodic driven system.  Instead, $S/S_P = \eta < 1$ for large $n$ and large $V_1$ as shown in the top left panel of Fig.\ \ref{hsdfig}. Interestingly, $\eta$ depends on the initial Fock state. This phenomenon was interpreted to be signature of fragmentation of the Hilbert space in the computational Fock basis in which case the state picked up always belongs to a definite fragment ({\it i.e} a Krylov subspace) whose HSD determines $\eta$. This feature is shown in the top right panel of Fig.\ \ref{hsdfig}.

In addition, a more experimentally accessible quantity, namely, the autocorrelation of fermion density $C_L(nT) = \langle \hat n_L(n T) \hat n_L(0)\rangle$ has also been studied for such systems. For ergodic system $C_L(nT) \to 0$ for $n \to \infty$ while for a system with Hilbert-space fragmentation, it always stays above the Mazur bound $C_b \sim 0.125$ \cite{majur1,hsf5}. It was found that at special frequencies and large drive amplitude, $C_L > C_b$. In contrast, at other frequencies or lower drive amplitudes, $C \to 0$ within a few drive cycles. This is shown in the bottom left panel Fig.\ \ref{hsdfig} for the model described by Eq.\ \ref{fermham} using open boundary conditions. 

Several other quantities of the driven fermion chain such as out-of-time correlators (OTOC) \cite{sg2} and equal time correlation functions \cite{sg1} show features different from an ergodic driven system. These constitute a signature of prethermal Hilbert-space fragmentation due to emergent approximate symmetries at special frequencies and high drive amplitude alluded to earlier. Some of these features, such as oscillations of OTOC indicating lack of information scrambling \cite{sg2} and oscillations of correlators staring from a frozen ${\mathbb Z}_2$ state \cite{sg1} have no analogue in fragmentation seen in equilibrium Hamiltonians.

The crossover from the fragmented to ergodic behavior at the special drive frequencies can be studied as a function of the drive amplitude. The fragmented behavior persists as long as $H_F^{(1)}$ controls the dynamics; this sets the prethermal timescale. This timescale turns out to be
exponentially sensitive to the drive amplitude at the special frequencies. Thus at low drive amplitudes, the prethermal regime shrinks to a few drive cycles while at large drive cycles it can be extremely large. In between, there is a sharp crossover between the two regimes as shown in the bottom right panel of Fig.\ \ref{hsdfig}; the crossover becomes flatter with increasing system size showing the finite-sized nature of this phenomenon.

Before ending this section, we note that most of the model Hamiltonians \cite{hsfrev}, studied in the context of equilibrium Hamiltonian, require imposition of additional constraints which may not be easy to implement; the prethermal version of this phenomenon does not have this requirement since the constraint is dynamically generated due to the effect of the drive. In this sense, it may have a better chance of realization in ultracold atom platforms \cite{rev12,hsfexp}.

\subsection{Floquet scars}
\label{qscars}

In classical physics, scars constitute trajectories of, for example, a moving billiard ball in a Buminovitch stadium, which do not cover the phase space completely. This represents non-ergodic behavior as known in classical dynamical systems for a long time \cite{scarcl}. Its implication in single-particle quantum mechanics has also been studied long back \cite{scarqm1}. In contrast quantum many-body scars are athermal, near mid-spectrum, eigenstates of non-integrable many-body Hamiltonians with low entanglement entropy. They have been discovered relatively recently \cite{exp3,scarrev}. Since then, quantum many-body scars have been widely investigated in context Rydberg atom platforms \cite{scarpxp1,scarpxp2,scarpxp3,scarpxp4,scarpxp5,scarpxp6,scarpxp7,scarpxp8,scarpxp9}, quantum spin models \cite{scarspin1,scarspin2,scarspin3,scarspin4,scarspin5,scarspin6,scarspin7,scarspin8,scarspin9,scarspin10,scarspin11,scarspin12}, interacting bosons \cite{scarbose1,scarbose2,scarbose3,scarbose4,scarbose5}, quantum link models \cite{scarlink1,scarlink2,scarlink3,scarlink4,scarlink5,scarlink6,scarlink7}, fermion models \cite{scarferm1}, open quantum systems \cite{scaropen1,scaropen2}, non-Hermitian models \cite{scarnh1}, and random network models \cite{mar1}.

The generic conditions required for the presence of such athermal scar eigenstates in the spectrum of a many-body quantum Hamiltonian is not completely understood. In a class of these models, their presence can be understood in terms of emergence of additional symmetries \cite{scarrev}; these lead to a quasiparticle picture for these states. In the Rydberg atom platforms, such emergent symmetries turn out to be approximate leading to decay of scar induced oscillations of correlation functions \cite{scarrev,scarpxp8}. In some other cases\cite{scarspin11,scarlink1,scarlink2,scarlink3,scarlink4,scarlink5,scarlink6}, the presence of such scars cannot be understood from such emergent symmetries; they do not admit a simple quasiparticle picture.

To understand the presence of such emergent symmetry, one may consider a simple model, namely a spin-one XY model in magnetic field discussed in this context in Ref.\ \onlinecite{scarrev}. The Hamiltonian of this model is given by
\begin{eqnarray}
H = -J \sum_{\langle ij\rangle} \left(S_i^+ S_j^- + {\rm h.c.} \right) - B_0 \sum_j S_j^z, \label{spinham1}
\end{eqnarray}
where $S_j^{\pm} = (S_j^x \pm i S_j^y)/2$ and $S_j^Z$ denotes components of $S=1$ spins on the site $j$ and $\langle i j\rangle$ indicates that $i$ and $j$ are nearest neighbors. The eigenstates of $S_j^Z$ are denoted as $|1\rangle$, $|0\rangle$, and $|-1\rangle$ with eigenvalues $1$, $0$ and $-1$ respectively (we have set $\hbar=1$). For large negative $B_0$, the ground state therefore is given by a Fock state which consists of down spins on all sites: $|G\rangle = |-1,-1,-1...-1\rangle$. It is to be noted that the first term of $H$ annihilates the ground state. 

The key observation for obtaining scar states in this model constitutes identifying a special class of excited eigenstates (bimagnons) that have similar properties. We note that the existence of such states is not guaranteed, unlike single magnon states, by charge ($S_z$) and total momentum  conservation \cite{scarrev}. These states can be constructed from $|G\rangle$ by repeated application of the operator
\begin{eqnarray}
J_+ &=&  \sum_{\ell} (-1)^{\ell} (S_\ell^{+})^2/2,  \label{scarop}
\end{eqnarray}
on the ground state and are given by \cite{scarrev}
\begin{eqnarray}
|n\rangle \sim (J_+)^n |G\rangle.   \label{scarst1}
\end{eqnarray}
There are $L+1$ such states and all of them are annihilated by action of the first term of $H$. Consequently, one finds a tower of eigenstates with energy $E_n= (n-2L) B_0$, where $L$ is the chain length in units of lattice spacing and $n=0,1...L$. The operators $J^+$, $J^- = (J^+)^{\dagger}$, and
\begin{eqnarray}
J_z &=& [J^+,J^-]/2 =\sum_{\ell} S_{\ell}^z/2,  \label{alg1}
\end{eqnarray}
form a spin-$L/2$ emergent ${\rm SU}(2)$ algebra. 

This emergent ${\rm SU}(2)$ symmetry is crucial for dynamical properties of the system. An initial state with large overlap with any of the bimagnon tower of states evolves under Hamiltonian dynamics within the bimagnon subspace due to this emergent symmetry. This feature leads to oscillatory behavior of spin-spin correlation functions; the long-time values of these correlation functions and the entanglement entropy differs from the prediction of ETH \cite{scarrev}. Notably, the
transition between the different scar (bimagnon) eigenstates are mediated by $J^{\pm}$. These local operators implement a spectrum generating algebra $[H, J^+]= \omega J^+$ where $\omega= E_n-E_{n-1}= B_0$ and thus
allows for a quasiparticle description \cite{scarrev}.

This simple picture of emergent symmetry, in its pristine form, does not hold for other, more complicated, spin models \cite{scarspin1,scarspin2,scarspin3,scarspin4,scarspin5,scarspin6,scarspin7,scarspin8,scarspin9,scarspin10,scarspin11} or Rydberg atom platforms \cite{scarpxp1,scarpxp2,scarpxp3,scarpxp4,scarpxp5,scarpxp6,scarpxp7,scarpxp8,scarpxp9,scarspin12}.
For example, the low-energy effective Hamiltonian of a Rydberg atom chain is given by \cite{exp4}
\begin{eqnarray}
H_{\rm Ryd} &=& \sum_j (-\Delta \hat n_j^z + \Omega \sigma_j^x)  +V_0 \sum_{j,j'} \frac{\hat n_j \hat n_{j'}}{|j-j'|^6}, \nonumber\\ \label{rydham}
\end{eqnarray}
where $\hat n_j$ denotes the density operator for the Rydberg excitation on site $j$ with $\sigma_j^z= 2\hat n_j-1$. Here $\Delta$ is the detuning parameter which can be tuned to
energetically favor Rydberg excitations, $\Omega$ denotes the coupling strength between the ground and Rydberg-excited states at site $j$, and $V_0$ is the van der Walls repulsion strength between
two excited Rydberg atoms. 

Such an interaction with large $V_0$ may exclude neighboring Rydberg excitations; this phenomenon is known as Rydberg blockade \cite{blref1}. Using this
property, it is possible to obtain $Z_n$ symmetry broken ground states of $H_{\rm Ryd}$ for large $V_0$ and $\Delta$ \cite{exp4}. In particular, the $Z_2$ symmetry broken state does not allow for
neighboring Rydberg excitations; in this limit, $H_{\rm ryd}$ may be described a constrained model given by \cite{ks1,ks2}
\begin{eqnarray}
H_{\rm spin} &=& \sum_j (-\lambda  \sigma_j^z + \Omega \tilde \sigma_j^x),  \label{fss}
\end{eqnarray}
where $\tilde \sigma_j^x = P_{j-1} \sigma_j^x P_{j+1}$, and $P_j= (1-\sigma_j^z)/2$ projects the spin at $j$ to $|\downarrow_j\rangle$. The PXP model is obtained from  $H_{\rm spin}$ for $\lambda =0$.

The PXP model has been shown to host quantum scar eigenstates \cite{scarpxp1,scarpxp2,scarpxp3,scarpxp4,scarpxp5,scarpxp6,scarpxp7,scarpxp8,scarpxp9}. An analysis of the model carried out in details in Refs.\ \onlinecite{scarpxp8} shows that there is no exact spectrum generating algebra \cite{scarpxp8,scarrev}. An identification
of $H_{\pm} = \sum_j (\tilde \sigma_{2j}^{\pm} + \tilde \sigma_{2j-1}^{\mp})$ leads to $[H_+,H_-]= S_z + {\mathcal O}_z $ where $S_z$ denotes the staggered magnetization, $\sum_j (-1)^j \sigma_j^z$, and ${\mathcal O}_z$ is given by  \cite{scarrev, scarpxp8}
\begin{eqnarray}
{\mathcal O}_z = \sum_j (-1)^j \sigma_{j-1}^z  \sigma_{j}^z  \sigma_{j+1}^z.
\end{eqnarray}
The presence of non-zero ${\mathcal O}_z$ indicates that the emergent ${\rm SU}(2)$ algebra is not closed; it constitutes the realization of an approximate ${\rm SU}(2)$ algebra \cite{scarpxp8}. 

The effect of such a non-closure can be seen in quench dynamics of the PXP chain starting from an initial $|{\mathbb Z}_2\rangle= |\uparrow, \downarrow, ... \uparrow, \downarrow\rangle$ Neel state. The initial oscillations of $\langle \hat n_j(T)\rangle$ decay to a steady state value where the decay rate depends on the amplitude of ${\mathcal O}_z$ term. The oscillatory behavior starting from $|{\mathbb Z}_2\rangle$ occurs because of its large overlap with the scar eigenstates of the model. The decay, in contrast, can be interpreted as the leakage to thermal sector due to the presence of the ${\mathcal O}_z$ terms. This indicates the presence of an approximate emergent ${\rm SU}(2)$ symmetry in these models. No such scar-induced oscillations are found if one starts from the vacuum $|0\rangle = |\downarrow, \downarrow, ...\downarrow \rangle$ state; such dynamics conforms to ETH predictions. The scar-induced ETH violation, due to such initial state dependence, is termed as {\it weak} and is to be contrasted with those due to integrability, MBL or Hilbert-space fragmentation. We also note that this symmetry based picture, even in the approximate sense, breaks down in complicated spin models \cite{scarspin11} and ${\rm U}(1)$ lattice gauge theories \cite{scarlink4,scarlink5} where no local operator with spectrum generating algebra can be constructed.

\begin{figure}
\rotatebox{0}{\includegraphics*[width=\linewidth]{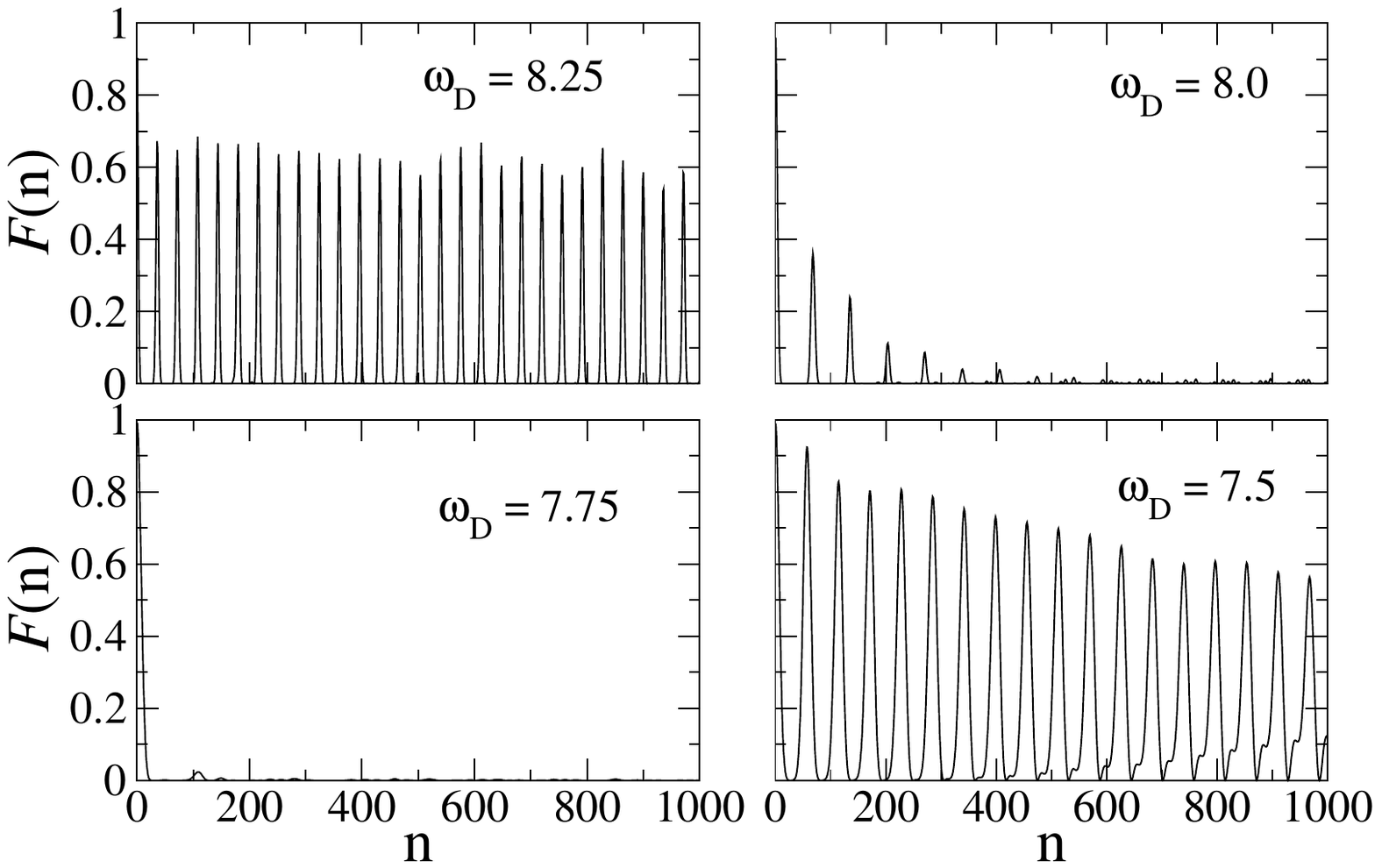}}
\caption{Plot of the fidelity ${\mathcal F}(n)= |\langle \psi(nT)|{\mathbb Z}_2\rangle|$ as
a function of $n$ showing periodic persistent revivals where the
dynamics is controlled by scars [top left and
bottom right panels] and fast decay with no subsequent revival in
their absence [bottom left panel]. An intermediate behavior
indicating crossover from coherent to thermal regime is shown in the
top right panel. For all plots, $\lambda_0=15 \Omega$ and  all figures are obtained from Ref.\ \onlinecite{bm1}.}
\label{figscar1}
\end{figure}
The Floquet version of this phenomenon has been studied in several works in the context of Rydberg atom platforms \cite{bm1,mituza1,bm2,lukinsc1,papic1}, fractonic models \cite{pretko1}, and trimerized Kagome lattice models \cite{liu1}.
For Rydberg atom platform, the drive was implemented by making the longitudinal magnetic field term $H_0= -\lambda \sigma_j^z$ in $H_{\rm spin}$ (Eq.\ \ref{fss}) time dependent; $\lambda$ was varied a function of time using a square-pulse \cite{bm1,bm2,mituza1}, periodic kicks \cite{lukinsc1}  or continuous cosine drive protocol \cite{papic1}. In each of these cases, scar eigenstates were found in the Floquet quasienergy spectrum of the
driven model; they were found to influence the dynamics of the model and leave their imprint on correlation functions and entanglement entropy.

For the square pulse protocol, it was found in Refs.\ \onlinecite{bm1,mituza1}, that the Floquet Hamiltonian admits scar states that have large overlap
$|{\mathbb Z}_2\rangle$ at high and intermediate frequency range except at special drive frequencies where the first order Floquet Hamiltonian vanishes. For the square-pulse protocol the first order Floquet Hamiltonian was found to be of the PXP form. It's perturbative analytical expression in the large drive amplitude regime is given by \cite{bm1}
\begin{eqnarray}
H_F^{(1)} &=& \Omega \frac{ \sin \lambda_0 T/4}{\lambda_0 T/4} \sum_j e^{-i \lambda_0 T/4} \tilde \sigma^+_j + {\rm h.c.} \label{fl1}
\end{eqnarray}
where $\lambda_0$ is the drive amplitude and $T$ is the time period of the drive. $H_F^{(1)}$ vanishes at special drive frequencies, $\omega_n^{\ast}$ for which $\lambda_0 T= 4 n \pi$ ($n \in Z$); away from these frequencies, at high drive amplitude, it controls the dynamics up to a large prethermal time scale. 

Since $H_F^{(1)}$ has the same form as $H_{\rm PXP}$, albeit with a rescaled $\Omega$ and an unimportant rotation in spin space, it's eigenspectrum hosts quantum scars. These scars lead to oscillatory dynamics of correlation functions, similar to the static case; such a behavior can be understood in the context of an emergent approximate $SU(2)$ symmetry as discussed above. Note that in the Floquet context this emergent symmetry is approximate both due to the non-closure of the ${\rm SU}(2)$ algebra and also from the fact that the higher order terms in the Floquet Hamiltonian do not respect such symmetry. The latter effect becomes particularly apparent at the special frequencies where $H_F^{(1)}$ vanishes; for these frequencies
one sees fast, ETH predicted, thermalization as can be seen in Fig.\ \ref{figscar1}. This allows for possibility of controlling quantum coherence using the drive frequency as a tuning parameter.

\begin{figure}
{\includegraphics*[width=0.98 \linewidth]{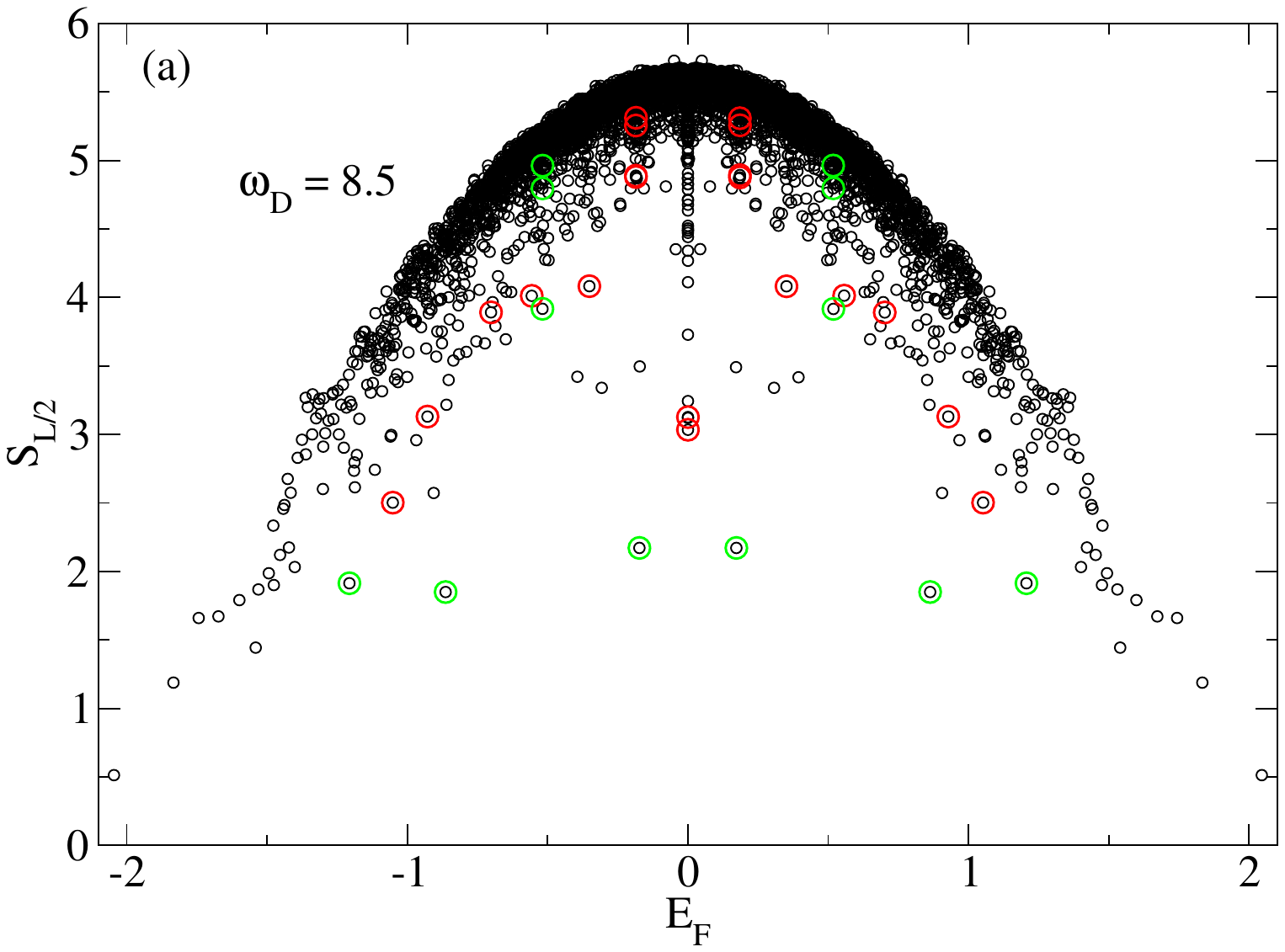}}
{\includegraphics*[width=0.98 \linewidth]{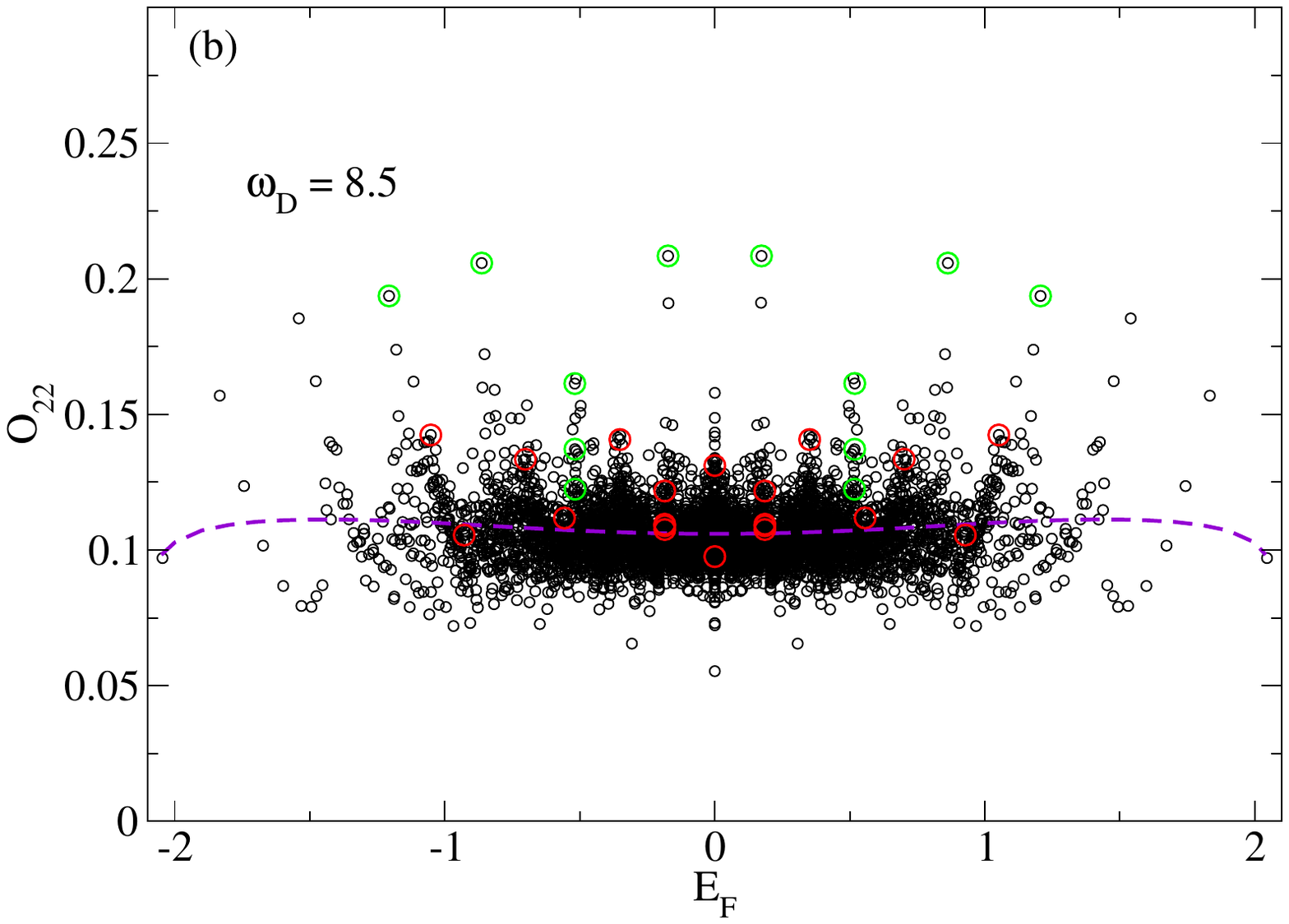}}
\caption{(a) Plot of $S_{L/2}$ for the eigenstates of $H_F$
for $L=26$ and $\hbar\omega_D/\Omega =8.5$ at $\lambda_0/\Omega=15$. The eigenstates with overlap $>0.01$
with $|0\rangle(|\mathbb{Z}_2\rangle)$ are shown using red (green) circles.
These states are distinct and coexist at this drive frequency.
(b) Plot of $O_{22}=\langle \hat n_2 \hat n_4\rangle$ as a function of Floquet eigenstate
quasienergies $E_F$ (in units of $\Omega$). The violet dashed line indicates the ETH
predicted value of $O_{22}$ at a temperature $T_0= E_F/k_B$. These figures, adapted from Ref.\ \onlinecite{bm2}, establishes 
coexistence of two different type of scar states in the driven PXP chain.}
\label{figscar2} \end{figure}

The Floquet scars found in Refs.\ \onlinecite{bm1,mituza1} are continually connected to the static scars of the PXP model in the high drive frequency limit. However, another class of scars which do not have any equilibrium analogue has also been found at intermediate frequencies \cite{bm2}. Unlike their counterparts which have large overlap with $|{\mathbb Z}_2\rangle$  Fock states, these show significant overlap with $|0\rangle$  state. They coexists with the ${\mathbb Z}_2$ scars at intermediate drive frequency range $8\le \hbar \omega_D/\Omega \le 10$. Such scars necessarily arise from the combined effect of higher order terms in the Floquet Hamiltonian. The first of such terms, derived using FPT in the high drive amplitude regime, is given by \cite{bm2}
\begin{eqnarray}
H_F^{(3)} &=& \sum_{j} \Big(A_0 [(\tilde \sigma_{j-1}^+ \tilde
\sigma_{j+1}^+ + \tilde \sigma_{j+1}^+ \tilde \sigma_{j-1}^+) \tilde
\sigma_j^- - 6 \tilde \sigma_j^+], \nonumber\\
&& +{\rm H.c.} \Big), \label{fl3} \\
A_0 &=& \Big[ e^{3i\lambda_0 T/(2\hbar)} +3 e^{i\lambda_0 T/(2\hbar)} (1+ i \lambda_0 T/\hbar) \nonumber\\
&& + 2 (1-3 e^{i\lambda_0 T/\hbar}) \Big] \frac{\Omega^3 e^{-i\lambda_0 T/\hbar}}{3 i \lambda_0^3 T/\hbar}.
\end{eqnarray}
It turns out that the eigenstates of $H_F^{(1)}+ H_F^{(3)}$ host additional scars which has large overlap with $|0\rangle$; these therefore do not have any equilibrium analogue. Their coexistence with the ${\mathbb Z}_2$ scar state is shown in Fig.\ \ref{figscar2}; the corresponding oscillatory dynamics of correlation function $\langle \hat n_j \hat n_{j+2} \rangle$ starting from the $|0\rangle$ state due to such scars has been shown in
Ref.\ \onlinecite{bm2}. The existence of these Floquet scars does not seem to be attributable to any emergent symmetry of the Floquet Hamiltonian.

More recently, a period-kick protocol has been studied in this context \cite{lukinsc1}. It has been shown that such a drive can induce, at specific value of the kick amplitude and frequency, a subharmonic response which can be interpreted as a signature of prethermal $Z_2$ time crystal. The difference of this phenomenon with its more standard counterpart (see Sec.\ \ref{tcr} for a detailed discussion) is that such a response occurs only when the starting state is a Neel state. The corresponding oscillations, which occur between the Neel state and its time-reversed partner, are controlled by $Z_2$ scars. These scar states share the same approximate emergent symmetry discussed earlier. In contrast, scar induced oscillations for both the Neel and the vacuum initial states have also been found for a cosine drive protocol in the non-perturbative regime \cite{papic1}. As in Ref.\ \onlinecite{bm2}, the origin of these scars can not be understood from an emergent symmetry principle.

\section{Prethermal time crystals}
\label{tcr}

The idea of a time crystalline phase of matter which breaks time-translational symmetry was first proposed in the context of equilibrium Hamiltonians in Ref.\ \cite{wil1}. However, it was later shown that such a phase can not exist \cite{bruno1,noz1,masaki1} in equilibrium. As noted in Ref.\ \onlinecite{masaki1}, such a phase can be diagnosed by an unequal-time correlation function of local operators $\hat A$ and $\hat B$: $C({\bf r-r'},t)=\langle \hat A({\bf r},t) \hat B({\bf r'},0)\rangle$, where the expectation is taken with respect to a many-body ground state. The presence of a time crystalline phase then constitutes a time-dependent long-time value of $C$ in the thermodynamic limit: 
\begin{eqnarray} 
\lim_{V \to \infty, |{\bf r}-{\bf r'}|\to\infty} C = f(t), \label{eqcond}
\end{eqnarray}
where $f(t)$ is a periodic function of time. Ref.\ \onlinecite{masaki1} explicitly showed that for local operators $\hat A$ and $\hat B$, $C$ can at best be a constant proving impossibility of such a phase in an equilibrium setting.

It was subsequently found that such phases can be realized in periodically driven quantum matter \cite{tcrev1,tcrev2,tcrev3,tcrev4,tcrev5,tcpap1,tcpap2,tcpap3}; they are dubbed as Floquet time crystals. A central problem encountered in realizing such phases constitutes drive-induced heating; Floquet systems tend to undergo fast thermalization to a featureless infinite temperature steady state \cite{rev7}. 

Several proposals for circumventing this problem were put forth. The first of them constitutes applying strong disorder leading to realization of a many-body localized phase which
evades such thermalization \cite{huserev1,ponte1,laza1,aba1}; this allows one to stabilize a discrete time crystalline phase \cite{tcrev1,tcrev2,tcrev3,tcrev4,tcrev5,tcpap1,tcpap3,sondhipap1}. The second is to use a domain-wall confinement mediated obstruction of rapid heating in an Ising chain \cite{domain1}. The third, which we shall be mostly interested in, is to consider a large drive frequency which suppresses energy absorption leading to a large prethermal regime where a time crystalline phase may be stabilized \cite{tcrev1,tcrev2,tcrev3,tcrev4,tcrev5,tcpap2,luitz1,sheng1,machado1,gam1}. This phenomenon has been experimentally investigated using several platforms such as ion traps \cite{tcexp1,tcexp2,tcexp3}, quantum circuits \cite{tcexp4}, NMR systems \cite{tcexp5,tcexp6}, superconducting qubits \cite{tcexp7,tcexp8} and quantum simulators\cite{tcexp9}. It has also been theoretically studied in the context of open quantum systems \cite{tcueda1,tcdem1,tcdsarma1}.

The key difference of such a prethermal time crystal with other prethermal phenomena discussed so far in this review involves the central role played by micromotion in realization of time crystals. To understand this, we consider a generic unitary evolution operator $U(T,0)$ corresponding to $H= H_0(t) + H_1(t)$. In the interaction picture, $U(T,0)$ an be written as 
\begin{eqnarray}
U(T,0) &=& T_t e^{-i \int_0^{T} H_1^{\rm int} (t') dt'/\hbar}, \nonumber\\
H_1^{\rm int}(t) &=& U_0^{\dagger}(t,0) H_1(t) U_0(t,0),
\label{uniteq1}
\end{eqnarray}
where $U_0(t,0) =T_t \exp[-i \int_0^t H_0(t') dt']$ such that $U_0(mT,0)=I$ for $m \in Z$ and $m>1$. It was shown in Ref.\ \onlinecite{tcpap2} for such a system there always exists a time-independent local unitary rotation ${\mathcal U}$. In the rotated frame, the evolution operator in the prethermal regime, is given by
\begin{eqnarray}
\tilde U(T,0) &=&  {\mathcal U}^{\dagger} U(T,0) {\mathcal U} \nonumber\\
&=& U_0(T,0) e^{-i H_F^{(1)}  T/\hbar} = X e^{-i H_F^{(1)}  T/\hbar}.
\end{eqnarray}
Here $X^m=I$,  $[X,H_F^{(1)} ]=0$, and $H_F^{(1)}$ is the perturbative first-order Floquet Hamiltonian. Note that $m=1$ reproduces the standard results discussed in earlier sections; in this case, the stroboscopic dynamics is completely controlled by $H_F^{(1)} $ which, being local in the prethermal regime, can not exhibit time crystalline behavior \cite{tcrev4,masaki1}. The commutation of $X$ with $H_F^{(1)} $ is most simply understood when $m=2$ ($X^2=I$); in this case, the requirement $\tilde U(T,0)= \tilde U(2T,T)$ leads to the relation \cite{tcrev4}
\begin{eqnarray}
X^{-1} e^{-i H_F^{(1)}  T/\hbar} X = e^{-i H_F^{(1)}  T/\hbar} \label{commutation}
\end{eqnarray}
from which the commutation $[X, H_F^{(1)} ]=0$ follows (for a more rigorous proof, see Ref.\ \onlinecite{tcpap2}). We note here that for $m\ne 1$, although $H_F^{(1)} $ is local in the prethermal regime, $\tilde U$ is not a local operator. Thus the arguments of Ref.\ \onlinecite{masaki1} can be circumvented in this case. 

The commutation of $H_F^{(1)} $ and $X$, in the rotated frame, indicates the presence of an additional approximate emergent internal symmetry of $H_F^{(1)} $ which leads to time crystalline behavior.  Such a symmetry was clearly absent in the microscopic Hamiltonian $H(t)$. Due to this internal symmetry, $H_F^{(1)} $ can be thought to have $m$ symmetry sectors. In a frame which moves with $X$, the operators there evolve under $H_F^{(1)} $ within one such symmetry sector. 

A rotation back to the lab frame therefore induces oscillations of operators with period $mT$ leading to a subharmonic response at $\omega= \omega_D/m$. These oscillations persist up to a prethermal time $t^{\ast}$ beyond which higher order terms in the Floquet Hamiltonian spoils the internal symmetry leading to melting of the time crystal. For the high drive frequency regime, $t^{\ast} \sim \exp[\omega_D]$ and the subharmonic oscillations persists for an exponentially long timescale.

To understand this phenomenon a bit more, we now consider a simple example which corresponds to a driven Ising chain with \cite{tcpap2}
\begin{eqnarray}
H_1(t) &=& -J \sum_{\langle i j\rangle} \sigma_i^z \sigma_j^z - h_z \sum_j \sigma_j^z, \nonumber\\
H_0(t) &=& h_x(t) \sum_j \sigma_j^x. \label{tceg1}
\end{eqnarray}
The simplest realization of time crystalline behavior can be obtained by  choosing a periodic kicked protocol at times $t= kT$, where $k \in Z$,  yielding \cite{tcpap2}
\begin{eqnarray}
h_x(t) = \pi \hbar \sum_k \delta(t- kT)/2. \label{transf}
\end{eqnarray}
This corresponds to $U_0(2T,0)= X^2= I$, where $X=\prod_j \sigma_j^x$ is a manifestly non-local operator. The corresponding $H_F$ can be computed within Floquet perturbation theory. The first order term, which controls the dynamics in the prethermal phase, is given by $H_F^{(1)}  = -J \sum_{\langle i j\rangle} \sigma_i^z \sigma_j^z$. This term commutes with $X$. The stroboscopic evolution operator is therefore given by (for this example ${\mathcal U}=I$ \cite{machado1})
\begin{eqnarray}
U(T,0) &=& \left(\prod_j \sigma_j^x\right) e^{i T J \sum_{\langle i j\rangle} \sigma_i^z \sigma_j^z/\hbar}. \label{evol}
\end{eqnarray}

The wavefunction of the Ising chain after $n$ cycles of the drive, where $n T\le t^{\ast}$ is therefore given by $ |\psi(nT)\rangle = U(nT,0)|\psi(0)\rangle$, where $|\psi(0)\rangle$ corresponds to a FM ground state
which leads to
\begin{eqnarray}
M(nT) &=& \frac{1}{L} \sum_j \langle \psi(nT)| \sigma_j^z|\psi(nT)\rangle \nonumber\\
&=& (-1)^n M(0), \label{respre}
\end{eqnarray}
indicating a subharmonic response in the prethermal regime. 

Interestingly the eigenstates $|E_{\pm}\rangle$ of $H_F^{(1)} $ with energy $E_z$
are macroscopic cat states. These are superposition of many-body Fock states given by $|z\rangle$
and its time-reversed partner $|{\bar z} \rangle$ with $|E_{\pm}\rangle = (|z\rangle \pm |{\bar z}\rangle)/\sqrt{2}$. These states satisfy $X|E_{\pm}\rangle = \pm |E_{\pm}\rangle$; in terms of these states one can write
\begin{eqnarray}
U(T,0) = e^{-i E_z T/\hbar} (|E_+\rangle \langle E_+| - |E_-\rangle \langle E_-|).\label{catrep}
\end{eqnarray}
The subharmonic response can be easily worked out from such a description. We note that in the present scenario, it is easy to see from these considerations that an initial state with $M(0)=0$ will fail to show any oscillations; one requires an initial macroscopic polarization. This feature distinguishes prethermal time crystals from their MBL counterparts which show stable oscillations irrespective of the chosen initial state \cite{tcrev2}.

\section{Discussion}
\label{diss}
In this brief review, we have discussed several prethermal phenomena in Floquet quantum systems. The key feature behind most of these
constitutes an additional approximate emergent symmetry which allows for the realization such phenomena. In all cases, such emergent
symmetry constitutes commutation of an operator ${\hat O}$ with a first-order Floquet Hamiltonian $H_F^{(1)}$ computed using some approximation scheme
in either high drive frequency or high drive amplitude regime. No such commutation exists for higher order terms in the Floquet Hamiltonian. This
feature essentially leads to prethermal nature of such phases; they are destroyed at long times when these higher order terms become important.
An exact computation of the prethermal timescale $t^{\ast}$ beyond which drive-induced heating takes over is a complicated problem. However, in the 
high drive frequency regime, $t^{\ast}$ can be estimated; it turns out to vary exponentially with the drive frequency. Such a large $t^{\ast}$ makes these 
prethermal phenomena experimentally relevant.

The most notable exception to this principle, for the problems addressed in this review, constitutes prethermal quantum scars. Some of these Floquet scars can be mapped to their equilibrium counterparts and can be understood in terms of emergent symmetries \cite{bm1,mituza1}. However, there is another class of such scars which have no obvious equilibrium analogue. It seems to be difficult to understand their existence from the point of view of an emergent symmetry \cite{bm2,papic1}. This difficulty partly stems from either the fact that they are not eigenstates of $H_F^{(1)}$ and necessarily require contribution from higher order terms in $H_F$ \cite{bm2} or from the fact that they exist in a non-perturbative regime where analytic computation of $H_F$ is difficult \cite{papic1}.

Another important aspect which is much less studied is the stability of the prethermal Floquet physics under experimentally relevant perturbations. Several phenomena such as quantum scars, Hilbert space fragmentation, and time crystalline order are comparatively new. Their analysis, so far,  is mostly confined to few model Hamiltonians and fine tuned regimes. The stability of such phenomena against disorder and other symmetry breaking or restoring perturbations will certainly be worth exploring. Similarly, an analytically tractable model for Hilbert-space fragmentation, similar to the AKLT model of exact quantum scars, will enhance the understanding these phenomena to a great extent and will be of interest.

It will be interesting to see these prethermal Floquet phases in higher dimensional system as they can be easily constructed in cold atom platforms. For example, Ref.\ \onlinecite{scar2d1} found exponentially many QMBS in 2D PXP model. It will be interesting to explore whether the Floquet version of this model or its generalization can also stabilize such scars \cite{scar2d2} in higher dimension and in different lattice geometries. Similar open question remain in the case of strong Hilbert space fragmentation. Recently \cite{as1} studied a 2D ring exchange spin model in the presence of subsystem symmetries and found signatures of strong Hilbert space fragmentation. Similar signatures were also found in Refs.\ \onlinecite{frank2,sid1} in other models. It will be interesting to investigate if there are prethermal version of such higher-dimensional Hilbert-space fragmentation in periodically driven models.

A relatively less explored related area of research constitutes understanding the effect of external bath on these phenomena. Some work has been done in this direction for time crystals \cite{tcueda1,tcdem1,tcdsarma1} and quantum scars in context of non-driven systems \cite{scaropen1,scaropen2}. However, for the latter class of problems, the existence or properties of Floquet scars have not been investigated in details. A similar generalization can be made for systems with prethermal Hilbert space fragmentation. Such open or non-Hermitian systems could be useful in preventing heating through bath or non-Hermitian term induced dissipation; they also allow for non-trivial steady states which can be controlled by the corresponding Lindbladians. These features may lead to possibility of engineering steady states with imprints of prethermal effects of closed systems. This issue has been studied for driven non-hermitian Ising chains \cite{tb1,tb2}; it certainly deserves a wider study in a broader context.

In conclusion, we have discussed several aspects of prethermal phases of Floquet quantum matter from the perspective of an approximate emergent symmetry which is crucial to their existence. Such systems exhibits dynamic freezing and localization and Hilbert space fragmentation. They also host Floquet scars and time crystalline phases. We have also discussed some open questions that can be subjects of future work in this area.

\section{Acknowledgments}

The authors thank Sayan Choudhury, Arnab Das, Somsubhra Ghosh, Bhaskar Mukherjee, Sourav Nandy, Roopayan Ghosh, Mainak Pal, Indranil Paul, Arnab Sen and Diptiman Sen 
for several discussions on related topics. KS thanks DST, India for support through SERB project JCB/2021/000030.

\end{document}